\documentclass[]{aa}

\usepackage[varg]{txfonts}
\usepackage{graphicx}
\usepackage{multirow}
\usepackage{ulem}
\bibpunct{(}{)}{;}{a}{}{,}

\usepackage{dashbox}

\begin{document}

\title{The HP2 Survey}
\subtitle{IV. The Pipe nebula: Effective dust temperatures in dense cores\thanks{HP2 stands for \textit{Herschel-Planck}-2MASS dust-optical-depth and column-density maps \citep[see][]{Lombardi2014,Zari2016,Lada2017}}\textsuperscript{,}\thanks{The reduced \textit{Herschel}-\textit{Planck} map, and the column-density and temperature maps are available at the CDS via...}}

\author{Birgit Hasenberger\inst{\ref{inst1}}
\and Marco Lombardi\inst{\ref{inst2}}
\and Jo\~{a}o Alves \inst{\ref{inst1}}
\and Jan Forbrich \inst{\ref{inst3}, \ref{inst4}}
\and Alvaro Hacar \inst{\ref{inst5}}
\and Charles J. Lada \inst{\ref{inst4}}}

\institute{Department for Astrophysics, University of Vienna, Türkenschanzstra\ss e 17, 1180 Vienna, Austria \label{inst1}
\and University of Milan, Department of Physics, via Celoria 16, 20133, Milan, Italy \label{inst2}
\and Centre for Astrophysics Research, University of Hertfordshire, Hatfield AL10 9AB, UK \label{inst3}
\and Harvard-Smithsonian Center for Astrophysics, 60 Garden Street, Cambridge, MA 02138, USA \label{inst4}
\and Leiden Observatory, Leiden University, PO Box 9513, NL-2300 RA Leiden, the Netherlands \label{inst5}}

\authorrunning{B. Hasenberger et al.}
\titlerunning{The HP2 Survey: IV. The Pipe nebula}

\abstract{Multi-wavelength observations in the sub-mm regime provide information on the distribution of both the dust column density and the effective dust temperature in molecular clouds. In this study, we created high-resolution and high-dynamic-range maps of the Pipe nebula region and explored the value of dust-temperature measurements in particular towards the dense cores embedded in the cloud. The maps are based on data from the \textit{Herschel} and \textit{Planck} satellites, and calibrated with a near-infrared extinction map based on 2MASS observations. We have considered a sample of previously defined cores and found that the majority of core regions contain at least one local temperature minimum. Moreover, we observed an anti-correlation between column density and temperature. The slope of this anti-correlation is dependent on the region boundaries and can be used as a metric to distinguish dense from diffuse areas in the cloud if systematic effects are addressed appropriately. Employing dust-temperature data thus allows us to draw conclusions on the thermodynamically dominant processes in this sample of cores: external heating by the interstellar radiation field and shielding by the surrounding medium. In addition, we have taken a first step towards a physically motivated core definition by recognising that the column-density-temperature anti-correlation is sensitive to the core boundaries. Dust-temperature maps therefore clearly contain valuable information about the physical state of the observed medium.}


\maketitle

\section{Introduction}

Dense cores are the direct predecessors of young stellar objects (YSOs), the cloud unit that will collapse to form a stellar system. Cores are likely to represent the end of a fragmentation hierarchy that characterises molecular clouds \citep[e.g.][]{Larson1981, Myers1983, Goodman1998}, and they embody the connection between molecular cloud structure and star formation. This connection poses questions which are among the most fundamental in the field, from the initial physical conditions of collapse to the origin of the Initial Mass Function \citep[e.g.][]{Motte1998, Alves2007, Lada2008, Andre2010, Konyves2015}. For that reason, it is critical to study entire molecular clouds complexes whose internal structure can be observed down to the scale of dense cores ($\simeq0.1$\,pc)  in order to understand the star-formation process.

Nearby molecular clouds are commonly mapped using observations of molecular line emission, dust extinction, and dust emission. Morphological analyses of these maps, for example for core extraction, typically focus on the structures visible in the respective density map. Multi-band observations of thermal dust emission additionally provide information on the dust temperature distribution of a region, which is not used frequently in observational core extraction studies \citep[see, however,][]{Marsh2015}. Due to the intricate systematic uncertainties associated with deriving dust temperatures, for example, noise and line-of-sight confusion \citep{Shetty2009, Shetty2009b}, it is not obvious a priori how closely dust temperatures are related to the physical state of the observed medium. In this work, we explore whether a statement on the thermodynamics of cores is possible based on dust-emission measurements of temperature.

The advent of the ESA \textit{Planck} and \textit{Herschel} missions has made it possible to use thermal dust emission at sub-mm wavelengths to map entire molecular cloud complexes at a relatively high resolution. In combination with high-fidelity 2MASS dust-extinction maps \citep{Lombardi2001, Lombardi2009}, the sub-mm data can provide directly calibrated column-density maps as well as temperature maps for most nearby molecular clouds. We began this series of papers with the Orion \citep{Lombardi2014}, Perseus \citep{Zari2016}, and California \citep{Lada2017} regions, in which we validated the technique. In this paper we present column-density and temperature maps of the \object{Pipe nebula}.

The Pipe nebula is located ${\sim}5^{\circ}$ to the north (in Galactic coordinates) of the Galactic centre at a distance of ${\sim}$130--145\,pc from the Sun \citep{Lombardi2006, AlvesFranco2007}. Although its total mass is ${\sim}10^4 M_{\odot}$ \citep{Onishi1999, Lombardi2006}, the Pipe nebula only hosts up to 21 YSOs \citep{Forbrich2009}, mostly in \object{Barnard~59} \citep{Brooke2007}. The resulting low star formation efficiency of $\sim$0.06\,\% and limited influence of stellar feedback imply that this region is ideal to study the early stages of star formation \citep{Alves2008}. Early CO observations of the Pipe nebula conducted by \cite{Onishi1999} revealed the filamentary structure of the cloud as well as a number of dense cores. Based on a near-infrared extinction map of the region \citep{Lombardi2006}, \cite{Alves2007}, and \cite{Rathborne2009} are able to define a core sample and draw a connection between the corresponding core mass function and the stellar initial mass function. \cite{Lada2008} find that the cores are in pressure equilibrium with their surroundings and postulate that their properties may be defined by only few parameters such as external pressure and temperature. Extensive molecular line observations towards cores investigated their chemical properties and evolution \citep[e.g.][]{Frau2012, Forbrich2014}. Polarimetry studies of the region highlight the influence of magnetic fields in the Pipe nebula \citep{AlvesFranco2008, Franco2010, Soler2016}. The properties of filaments in the north-western part of the Pipe nebula, including the actively star-forming B59, indicate that the filamentary structure of that region may be the result of a compression by the wind from a nearby OB association \citep{Peretto2012}. Due to their isolation and quiescent environment, the two cores \object{Barnard~68} and \object{FeSt~1-457}, both located in the Pipe nebula, are regarded as prototypical prestellar cores. Multiple studies investigate their characteristics, including density and temperature distributions \citep{Nielbock2012, Roy2014}, dust properties \citep{Ascenso2013, Forbrich2015}, kinematics \citep{Aguti2007}, and core evolution \citep{Burkert2009}.

\section{Data}
\label{sec:data}

To create maps of the Pipe nebula based on dust emission, we have used data obtained by the \textit{Herschel} \citep{Pilbratt2010} and \textit{Planck} \citep{Planck2011} satellites. \textit{Herschel} observed the Pipe nebula using the instruments PACS \citep{Poglitsch2010} and SPIRE \citep{Griffin2010} in the course of the Gould Belt survey \citep{Andre2010}, which aimed at studying nearby molecular cloud complexes.

Maps of thermal dust emission \citep{Planck2014} that were derived from \textit{Planck} data are used to achieve an absolute calibration of the \textit{Herschel} flux maps and to fill in regions without \textit{Herschel} coverage. The \textit{Planck} maps themselves are calibrated based on the solar dipole and by comparing measured and modelled flux densities from the planets Uranus and Neptune. To set the zero-point of the maps, they were correlated with HI emission in low-column-density regions and thus provide an absolute calibration reference \citep{Planck2014calibration}. We note the key steps of the \textit{Planck} dust model \citep{Planck2014} in the following. In order to model the dust emission traced by the \textit{Planck} instruments, a modified black body (MBB) is used to fit the spectral energy distribution. This model contains three free parameters; the optical depth, the effective temperature, and the spectral index of dust emissivity. The model suffers from degeneracies between parameters, most prominently between the latter two \citep[e.g.][]{Shetty2009, Planck2014}. By performing the fit in a two-step process, this degeneracy can be alleviated, however \citep{Planck2014}. First, the spectral index is determined on a lower resolution of 30$^\prime$, and then the optical depth and effective temperature are fitted on a 5$^\prime$ scale assuming a fixed spectral index based on the 30$^\prime$ map. The model is reported to fit the data within the associated uncertainties. In a more recent study, dust emission observed by the \textit{Planck}, IRAS, and WISE satellites is fitted by a physical dust model \citep{Planck2016}. However, the study reveals complications with the assumptions of the model and the resulting column density required an empirical renormalisation to match independent measurements. In order to avoid the complex systematics associated with a physical dust model and maintain the consistent use of a dust emission model throughout this paper series, we have adopted the \textit{Planck} maps based on an MBB fit. For the conversion of dust optical depth to column density, we used a dust extinction map as reference that is based on 2MASS \citep{Skrutskie2006} data and was created with the NICEST algorithm \citep{Lombardi2009}.

\section{Data reduction}
\label{sec:methods}

The data reduction is identical to the data reduction described in detail by \cite{Zari2016}. In the following, the essential steps of this procedure are summarised briefly.

\begin{figure*}
  \resizebox{\hsize}{!}{\includegraphics{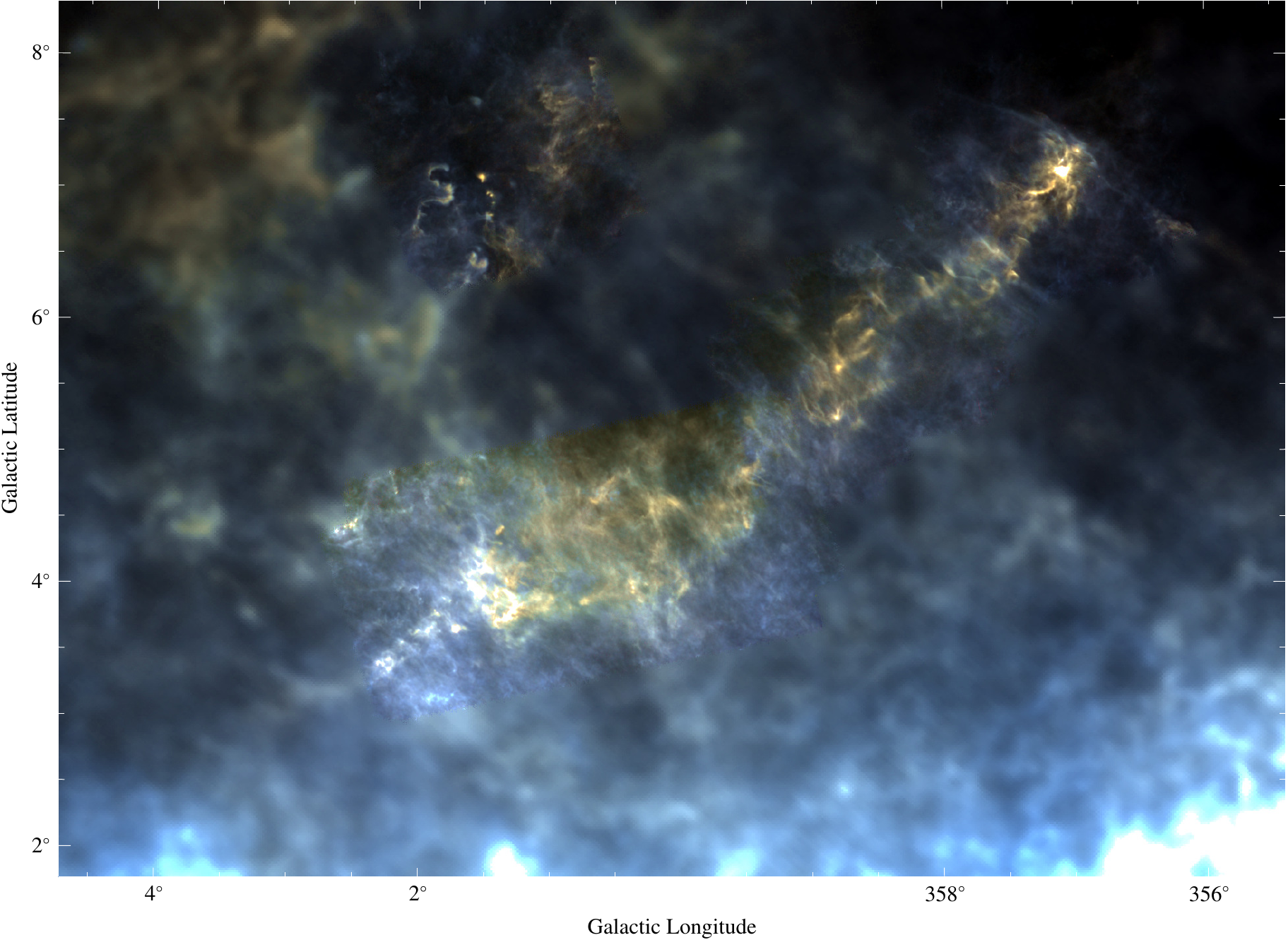}}
  \caption{Three-colour composite image of the intensity at the SPIRE bands, where blue, green, and red correspond to the bands centred at 250\,$\mu$m, 350\,$\mu$m, and 500\,$\mu$m, respectively. In regions not covered by \textit{Herschel}, the \textit{Planck} dust model parameters were used to predict the intensity at the SPIRE bands.}
  \label{fig:SPIREcomposite}
\end{figure*}

As a basis for our analysis, we produced Unimaps and HIPE level 2.5 data products for the PACS and SPIRE bands, respectively. The \textit{Herschel} flux values were calibrated to absolute values by comparison with \textit{Planck} data. To this end, the \textit{Planck} dust maps and the corresponding dust model were used to estimate the flux in each \textit{Herschel} band, which was compared to the \textit{Herschel} flux value. For each field and band, the relation was fitted with a linear function representing the absolute calibration. Figure~\ref{fig:SPIREcomposite} shows a composite image of the emission mapped by the three SPIRE bands in the Pipe nebula region. The smooth transitions between areas covered by \textit{Planck} only and \textit{Herschel} indicate that the calibration was successful for the majority of observation fields. In the central field, the \textit{Herschel} data show a gradient along Galactic latitude that is not observed with \textit{Planck}. We were therefore unable to achieve a satisfactory calibration over the entire central field, which causes a discontinuity along its northern edge. Throughout the data reduction and analysis, we monitored this region of the map closely and found that it generally does not present atypical properties. We therefore refrain from treating the central field separately in this work and state any significant deviations explicitly in the text.

Next, an MBB model analogous to the \textit{Planck} dust model (see Sect.~\ref{sec:data}) was fitted to the calibrated \textit{Herschel} data for the three SPIRE bands with central wavelengths at ${\sim}$250\,$\mu$m, 350\,$\mu$m and 500\,$\mu$m and one PACS band with its central wavelength at ${\sim}$160\,$\mu$m. For this step, all flux maps were degraded to the resolution of the SPIRE map at 500\,$\mu$m, namely 36$^{\prime \prime}$. The dust model can be written as
\begin{equation}
I_{\nu} \simeq B_{\nu}(T)\ \tau_{\nu},
\end{equation}
where $B_{\nu}(T)$ corresponds to Planck's law at the effective temperature~$T$ and~$\tau_{\nu}$ is the optical depth. The approximation is only valid when $\tau_{\nu} \ll 1$, which is fulfilled for thermal dust emission at sub-mm wavelengths \citep{Planck2014, Lombardi2014}. The wavelength dependency of~$\tau_{\nu}$ was assumed as
\begin{equation}
\tau_{\nu} = \tau_{\nu_0} \left(\frac{\nu}{\nu_0}\right)^{\beta},
\end{equation}
and the reference frequency $\nu_0$ was chosen to be 353\,GHz (850\,$\mu$m), the same as in the \textit{Planck} dust maps \citep{Planck2014}. We used a $\chi^2$ minimisation to perform the fit, taking both photometric and calibration errors into account. The optical depth and effective temperature were treated as free parameters, while the spectral index $\beta$ was set to the values found in the \textit{Planck} maps. We thus obtained maps of~$\tau_{850}$ and~$T$, along with maps of their corresponding errors. For simplicity, we denote the effective dust temperature as $T$ in this work, in contrast to the thermodynamic gas temperature $T_g$.

The $\beta$ map used here has a significantly lower resolution (30$^\prime$) than the \textit{Herschel} observations, meaning that any variation in $\beta$ on scales smaller than 30$^\prime$ is not considered. However, \citet{Forbrich2015} demonstrate that towards the cores FeSt 1-457 and B68 no significant variations of $\beta$ can be observed on scales between 36$^{\prime\prime}$ and 30$^\prime$. For the resolution of our HP2 maps, the spectral index based on the \textit{Planck} map is therefore likely sufficient to describe spatial changes in $\beta$ in this nearby cloud.

\begin{figure}
  \resizebox{\hsize}{!}{\includegraphics{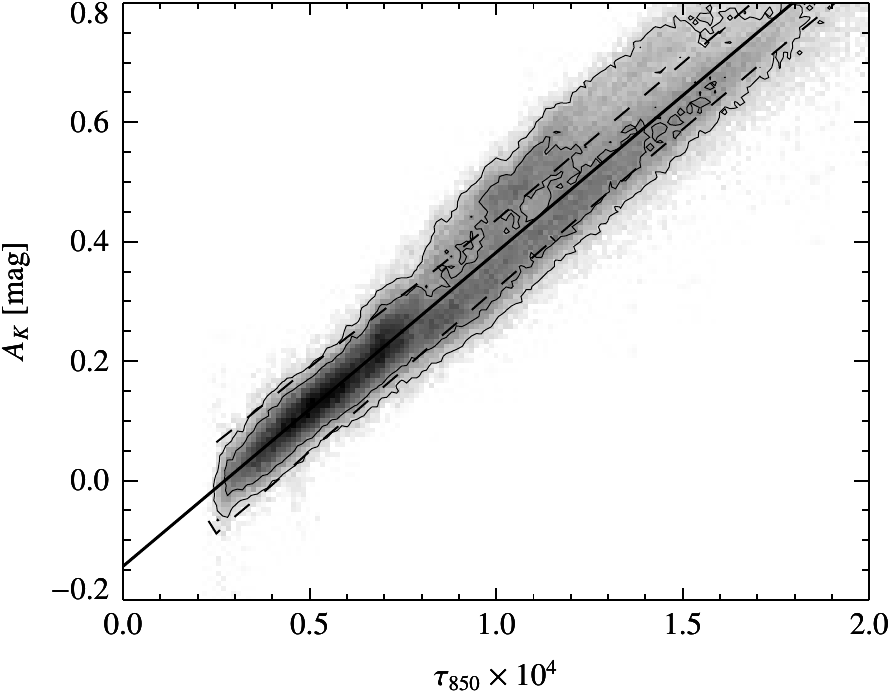}}
  \caption{Relation between NIR extinction values $A_K$ and optical depth derived from dust emission $\tau_{850}$. The solid line indicates the best-fitting linear function in the $\tau_{850}$ range shown, the dashed line the corresponding 3-$\sigma$ region. The black contours enclose 68\,\% and 95\,\% of the data points.}
  \label{fig:AKtau}
\end{figure}

In order to convert optical depth to column density expressed as an extinction value, the $\tau_{850}$ map was compared with an NIR extinction map of the Pipe nebula (see Sect.~\ref{sec:data}). Assuming a linear relationship in which
\begin{equation}
A_K = \gamma \tau_{850} + \delta
\end{equation} 
for optical depths $< 2\times10^{-4}$, we obtained the calibration factor $\gamma$ and intercept $\delta$ via $\chi^2$ minimisation (see Fig. \ref{fig:AKtau}):
\begin{align*}
\gamma &= \left(5256 \pm 6\right)\,\mathrm{mag},\\
\delta &= \left(-0.1436 \pm 0.0006\right)\,\mathrm{mag}.
\end{align*}
The given errors include formal fit errors only and do not take systematic uncertainties into account, for example those associated with the flux calibration or the choice of the fitting function.

An intercept significantly different from zero indicates that the flux calibration of \textit{Herschel} data is inaccurate, or that there is a shift in the extinction map, possibly caused by a suboptimal control field, or both. The smooth transitions from \textit{Herschel} to \textit{Planck}-only regions as well as the consistency between calibration parameters across most \textit{Herschel} fields (with the exception of the central field, see above), suggest that the flux calibration is not the main cause of the intercept. The search for an extinction-free control field is challenging due to the proximity of the Pipe nebula to the Galactic plane and the complex structures associated with this region. Indeed, the region selected as a control field in the north-west corner of the HP2 map is not entirely devoid of dust emission (see Fig.~\ref{fig:SPIREcomposite}). We therefore considered the intercept likely to be caused by extincted stars in the control field and did not include it in the conversion from optical depth to column density. 

The slope provides information on the properties of dust along the line of sight because of the connection between $\gamma$, the opacity at 850\,$\mu$m, and the K-band extinction coefficient \citep{Lombardi2014, Zari2016}. For the Orion, Perseus, and California molecular clouds slopes in the range ${\sim}2600 - 3900$\,mag were determined \citep{Lombardi2014, Zari2016, Lada2017}, while we found a significantly larger value for the Pipe nebula. This possibly indicates differences in the dust composition or grain size distribution. Several different dust models and their respective predictions for the value of $\gamma$ were compared with the results for the Orion molecular clouds \citep{Lombardi2014}. The considered models by \citet{Mathis1977}, \citet{Ossenkopf1994}, and \citet{Weingartner2001} produce $\gamma$ values too high for the Orion clouds, but comparable to the value derived for the Pipe nebula. Conversely, certain models by \citet{Ormel2011} fit well for Orion while they appear to be unable to explain the $\gamma$ value for the Pipe nebula. With the available data, it remains unclear whether the dust model assumptions represent actual differences in the dust properties between the cloud complexes, or whether the dust models require a revision so that they accommodate a larger range of $\gamma$ values for similar sets of dust properties. Clearly, an extensive study of the assumptions and predictions of various dust models in comparison with the results of dust emission observations is required to allow for a coherent statement on the variation of dust properties between different molecular clouds.

For optical depths beyond $2\times10^{-4}$, the relation between $\tau_{850}$ and $A_K$ is not approximated well with a linear function. Deviations from the linear relation can be explained by a lack of background stars at high column densities, which leads to inaccurate results in the extinction map. This phenomenon occurs in distinct regions that constitute ${\sim}$0.8\% of the total map area (see discussion in Sect.~\ref{subsec:compext} and Fig.~\ref{fig:compextmap}), e.g. the cores B59, B68, and FeSt 1-457 due to their high peak column density. To avoid an influence of this effect on the final maps, we assumed the linear correlation derived at lower extinctions for the conversion to column density.

\section{Column-density and temperature maps}

\begin{figure}
  \resizebox{\hsize}{!}{\includegraphics{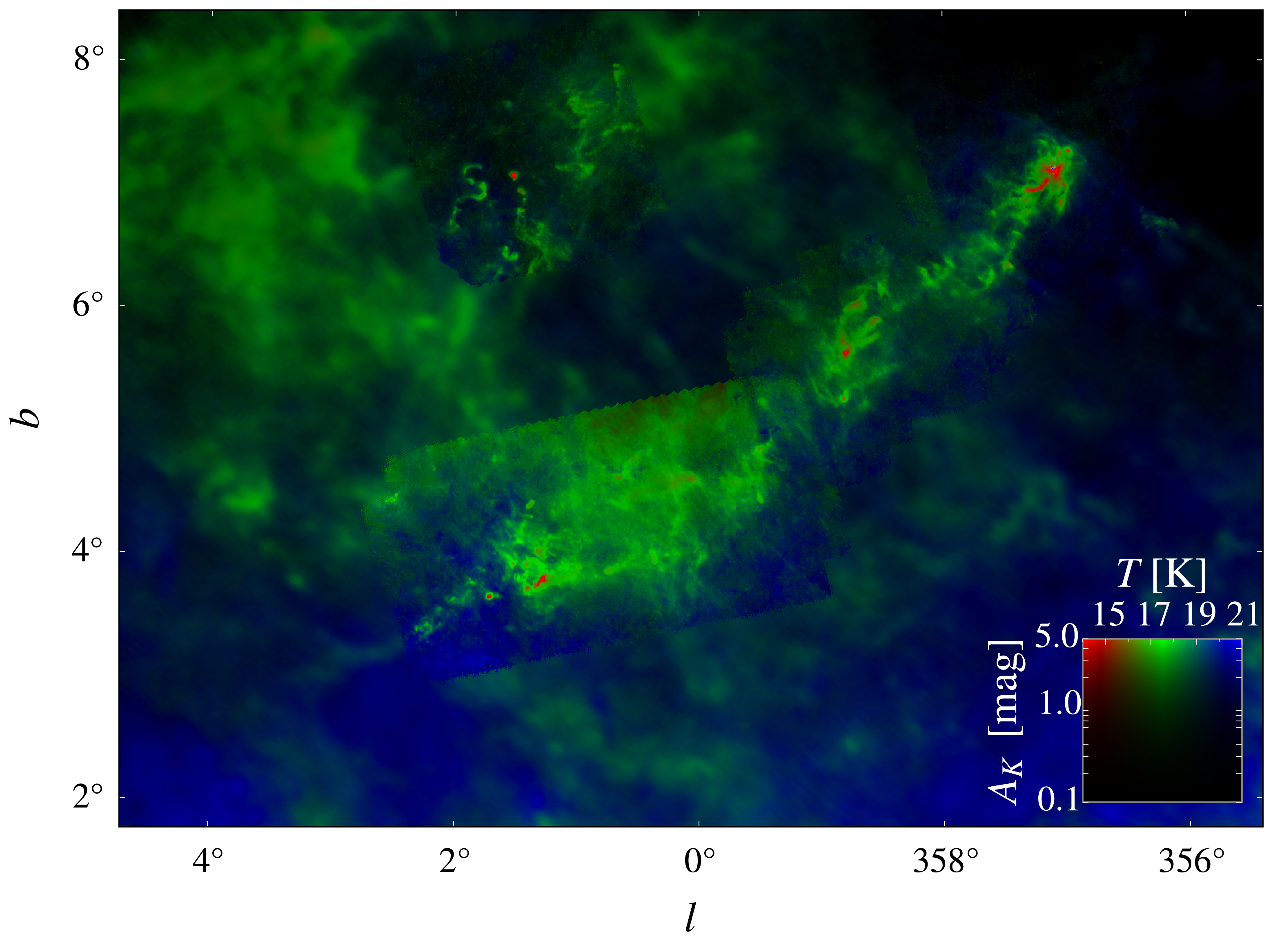}}
  \caption{Combined image of column density $A_K$ and effective temperature $T$ from the HP2 maps. Column density is shown as image brightness while effective temperature is shown as hue.}
  \label{fig:tauTmashup}
\end{figure}

\begin{figure}
    \centering
    \includegraphics[width=\hsize]{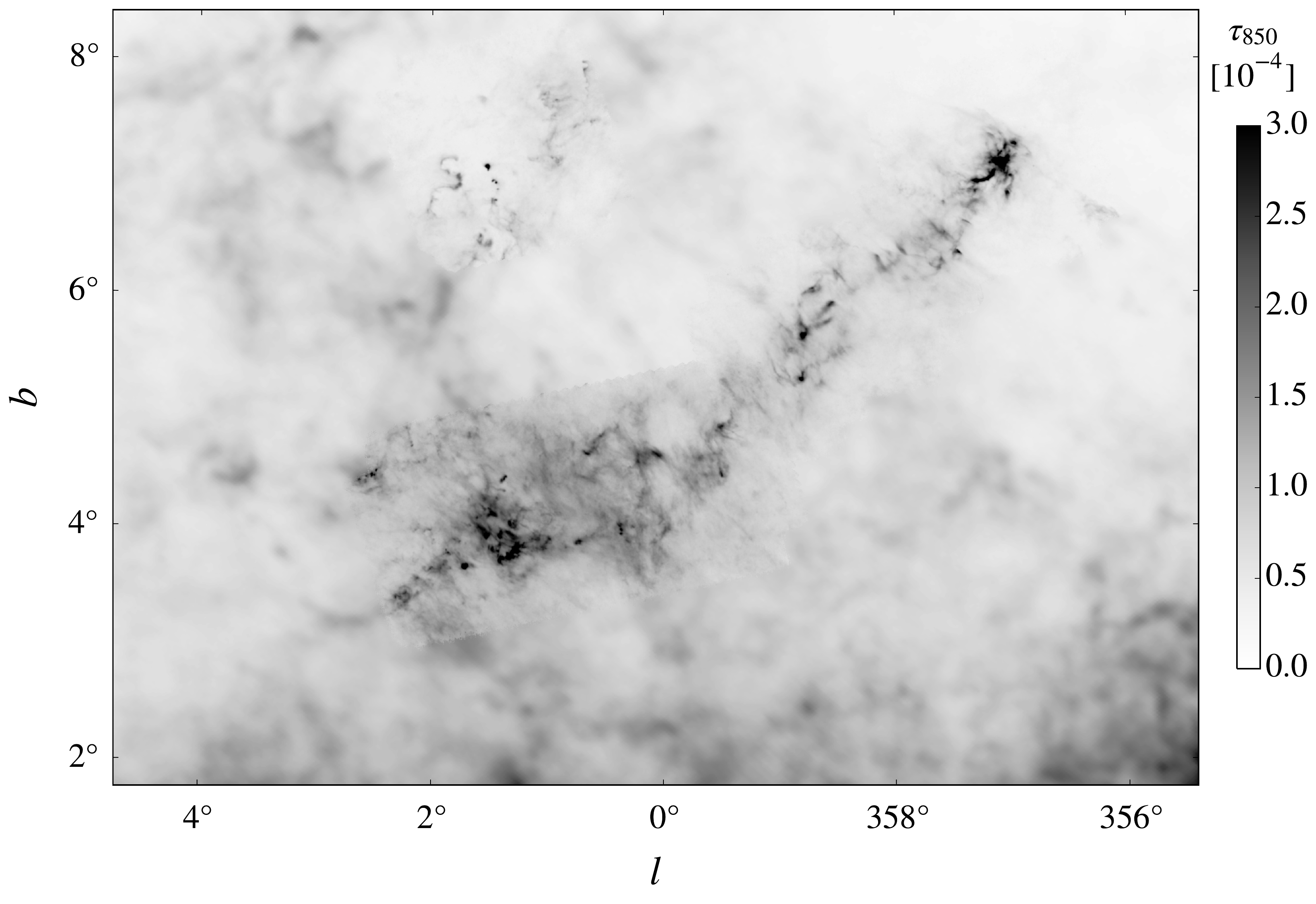}%
    \caption{Maps of the optical depth $\tau_{850}$ and the corresponding error $\sigma_{\tau_{850}}$ derived from dust emission, shown as different layers of this figure.}
    \label{fig:taumap}
\end{figure}

\begin{figure}
    \centering
    \includegraphics[width=\hsize]{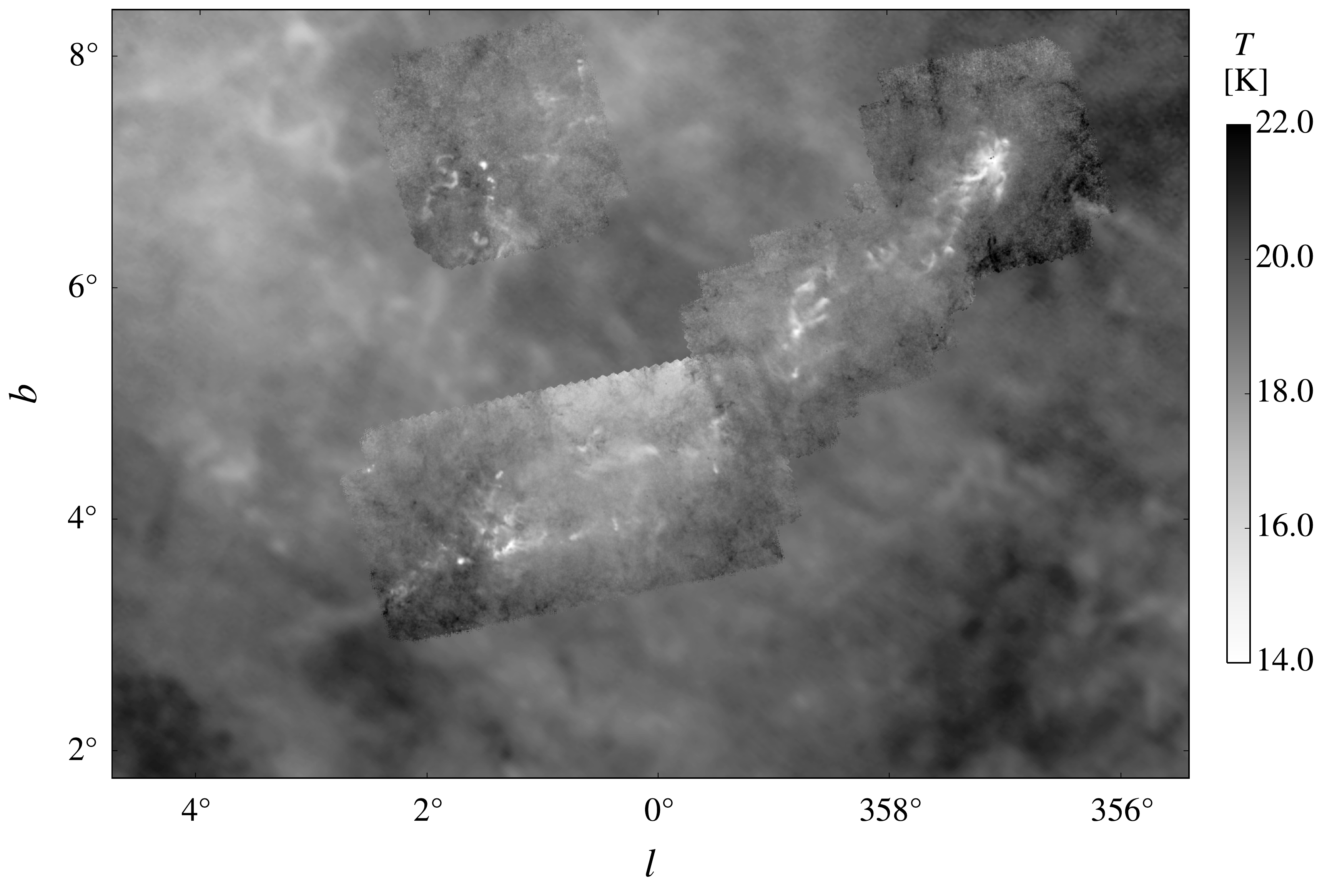}%
    \caption{Maps of the effective temperature $T$ and the corresponding error $\sigma_T$ derived from dust emission, shown as different layers of this figure.}
    \label{fig:Tmap}
\end{figure}

The final maps are shown as a combined image in Fig.~\ref{fig:tauTmashup} and as individual maps in Figs.~\ref{fig:taumap} and~\ref{fig:Tmap}. The resolution in the regions observed by \textit{Herschel} and only by \textit{Planck} is 36$^{\prime \prime}$ and 5$^\prime$, respectively. In optical depth, the map covers a range from ${\sim}2\times10^{-5} - 2\times10^{-3}$ with a mean error of ${\sim}3\times10^{-6}$, which translates to a range of ${\sim}0.1 - 12$\,mag and a mean error of ${\sim}0.01$\,mag in K-band extinction. The effective temperature map covers a range from ${\sim}13 - 24$\,K with a mean error of ${\sim}0.4$\,K. These values are derived from the entire map, including both areas observed by \textit{Herschel} and areas observed by \textit{Planck} only. The dynamic range of the entire map in $\tau_{850}$ and $T$ is set by the \textit{Herschel} data since those observations cover the densest parts of the cloud and have higher resolution. The mean errors are larger in the regions observed only by \textit{Planck} than in those with \textit{Herschel} data by factors of approximately three and five for optical depth and effective temperature, respectively.

\subsection{Global statistical analysis}

\begin{figure}
  \resizebox{\hsize}{!}{\includegraphics{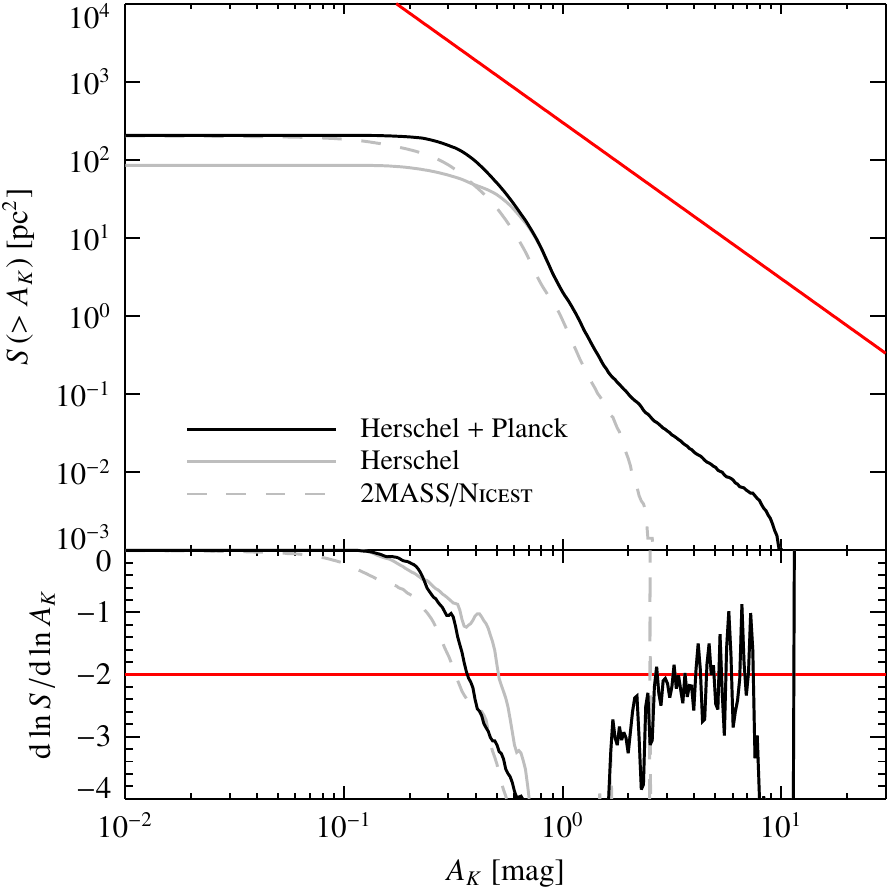}}
  \caption{\textit{Upper panel:} Integrated surface area $S(>A_K)$ as a function of extinction threshold $A_K$. \textit{Lower panel:} Derivative of the integrated surface area with respect to extinction, both converted to a logarithmic scale, as a function of extinction threshold. The black and grey solid lines are derived from the HP2 map, namely the entire mapped field and the regions covered only by \textit{Herschel}, respectively. The dashed grey line was calculated from the NIR extinction map. The red line corresponds to the toy model by \cite{Lombardi2014}, a power-law function with a slope of~$-2$.}
  \label{fig:Sdist}
\end{figure}

\begin{figure}
  \resizebox{\hsize}{!}{\includegraphics{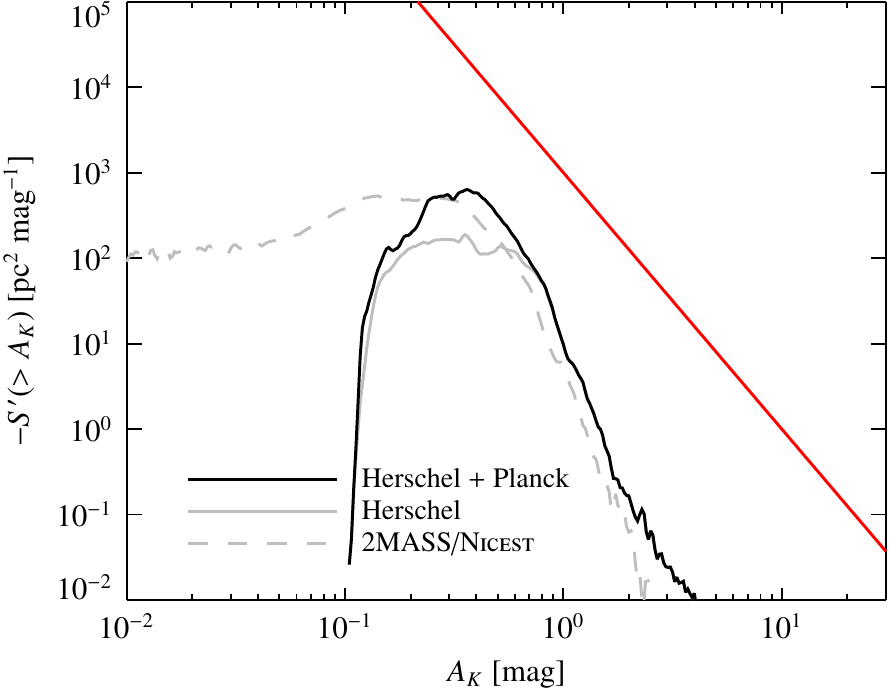}}
  \caption{Derivative of the integrated surface area with respect to extinction $S^{\prime}(>A_K)$ as a function of extinction threshold $A_K$. This function is proportional to the PDF of the region \citep{Lombardi2014}. The colour coding of the lines is as in Fig.~\ref{fig:Sdist}. The red line corresponds to the toy model by \cite{Lombardi2014}, a power-law function with a slope of~$-3$.}
  \label{fig:Sprimedist}
\end{figure}

Statistical distributions such as the surface area function~$S$ or the probability density function (PDF) enable us to make statements on the global structure of the mapped region. The surface area function is given by the area above a (variable) extinction threshold. The shape of~$S$ has been proposed to influence the level of star-forming activity in a cloud \citep{Lada2013}. Assuming that the radial density profile of a singular isothermal sphere applies to the dense parts of molecular clouds, the surface area function is proportional to~$A_K^{-2}$ \citep{Lombardi2014}. For the Orion and Perseus molecular clouds, this toy model describes the shape of~$S$ well over approximately two orders of magnitude in $A_K$ \citep{Lombardi2014, Zari2016}. This is not the case, however, for the Pipe nebula (see Fig.~\ref{fig:Sdist}). Similar conclusions can be drawn from the PDF (Fig.~\ref{fig:Sprimedist}), the mass-extinction relation (Fig.~\ref{fig:Mdist}), and its derivative (Fig.~\ref{fig:Mprimedist}) which have a more complex shape than those derived for Orion and Perseus. This could indicate that the Pipe nebula itself has a different internal structure incompatible with the assumptions of the toy model, or that other objects along the line of sight confuse our picture. Using data from the NIR extinction map of the Pipe nebula, \cite{Lombardi2006} also find a PDF with a complex shape and relate peaks in the distribution to background clouds. Since the region is located close to the Galactic plane, the background is highly spatially variable and thus complex to model, generally making it difficult to obtain a reliable PDF of the cloud. After estimating and removing background contributions \citep{Lombardi2015}, the PDF slope\footnotemark[1] of the Pipe nebula was found to be ${\sim}-4$; steeper than all other observed clouds except Polaris \citep{Lombardi2015} and similar to the California cloud \citep{Lada2017}. These results are in agreement with the hypothesis that steep PDF slopes are correlated with low star-formation activity \citep{Lada2013,Stutz2015}, since both the Pipe nebula and the California cloud exhibit low star-formation rates compared to clouds of similar total mass \citep[see][]{Lada2010}.

\footnotetext[1]{The power law functions given here differ from those given by \citet{Lombardi2015} by a multiplicative factor proportional to $A_K$ since we apply linear as opposed to logarithmic bins. This translates to a difference of one in the derived slopes.}

\subsection{Comparison with NIR extinction map}
\label{subsec:compext}

We assessed the consistency between the HP2 column-density map and the extinction map by comparing them on a pixel-by-pixel basis, and by comparing properties of a core sample defined previously in the literature. Although the HP2 optical-depth map was calibrated using the extinction map, this comparison is a useful tool to evaluate whether systematic differences occur. The calibration serves the purpose of converting optical depth to K-band extinction via a single factor. In the ideal case, the two maps would then be equivalent and deviations occur in a random fashion with respect to position, column density, and effective temperature. Systematic effects inherent to the data and methods used to create the maps could, however, be apparent despite the calibration.

In the following, we have used a previously defined core sample (henceforth referred to as R09 cores) by \citet{Rathborne2009}, where core boundaries are based on the application of the \textit{clumpfind} algorithm \citep{Williams1994} on a NIR extinction map \citep{Lombardi2006}. Molecular line observations of C${^{18}}$O were used to reject extinction peaks and to examine the internal structure of regions outlined by \textit{clumpfind}. Differences between the extinction map presented by \cite{Lombardi2006} and the map used in this work result from the use of the NICER and NICEST algorithm, respectively, and are generally minor.

\begin{figure}
  \resizebox{\hsize}{!}{\includegraphics{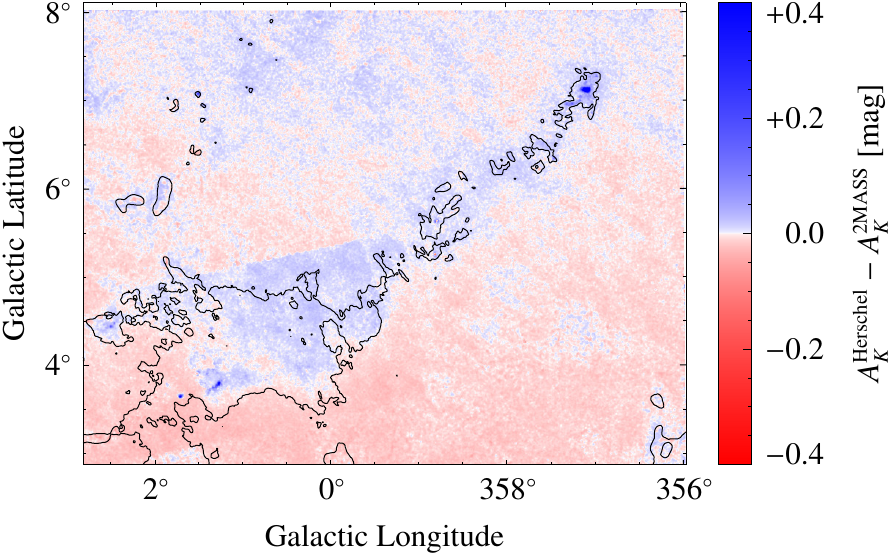}}
  \caption{Map of the difference between extinction derived from dust emission $A_K^{\mathrm{Herschel}}$ and from NIR dust extinction $A_K^{\mathrm{2MASS}}$. The contour represents the $A_K^{\mathrm{Herschel}}=0.5$\,mag level.}
  \label{fig:compextmap}
\end{figure}

First, the extinction map was reprojected to match the pixel scale and projection of the HP2 maps. Next, the maps were compared directly on a pixel-by-pixel basis, as shown in Fig.~\ref{fig:compextmap}. The intercept~$\delta$ (see Sect.~\ref{sec:methods}) was taken into account in this comparison to avoid a constant offset between the two extinction values. Similar to the results for the Orion, Perseus, and California molecular clouds \citep{Lombardi2014, Zari2016, Lada2017}, we found that the emission map gives higher $A_K$ values than the extinction map in regions of high column density, possibly due to the higher resolution and the smaller number of background stars observed in those areas. Alternatively, this phenomenon can be explained by an increased dust opacity as the result of changes in dust properties towards dense regions \citep[see, e.g.][]{Ysard2012, Planck2014, Juvela2015}.

\begin{figure}
  \resizebox{\hsize}{!}{\includegraphics{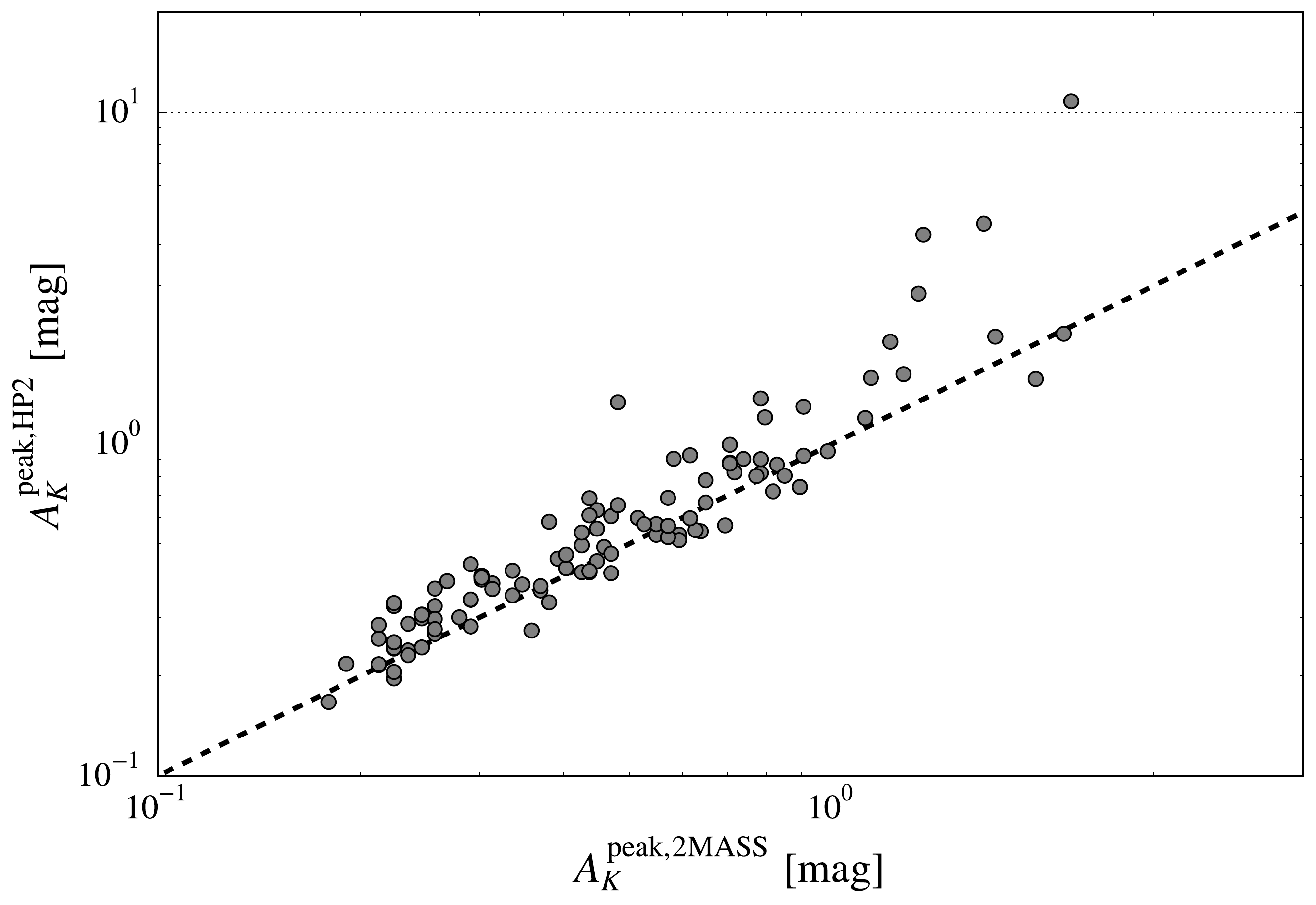}}
  \caption{Comparison of core peak column densities derived by \cite{Rathborne2009} $A_K^{\mathrm{peak, 2MASS}}$ and from the HP2 map $A_K^{\mathrm{peak, HP2}}$. The dashed line indicates equality.}
  \label{fig:peakAKcomp}
\end{figure}

In order to test these interpretations, we examined an NIR extinction map obtained for two cores in the Pipe nebula, FeSt 1-457 and B68, with a resolution higher than our 2MASS map by a factor of $\simeq6$ \citep{Forbrich2015}. In both cores, the relation between column densities derived from \textit{Herschel} data and higher-resolution NIR extinction is linear up to extinction values of $A_K\simeq3$\,mag. This value corresponds to the peak extinction in B68. In the FeSt 1-457 core, however, the $\tau_{850}$-$A_K$ relation deviates from the linear trend towards higher dust-emission column densities for $A_K\gtrsim3$\,mag, which is interpreted as an increase in dust opacity. Using only 2MASS data, we found that deviations from a linear $\tau_{850}$-$A_K$ relation occur at lower extinction values of $\simeq1$\,mag. Thus, differences between the 2MASS extinction and the HP2 map in these cores are caused predominantly by the lack of background stars for $A_K\lesssim3$\,mag. Given that only few cores in the R09 sample reach such extinction values (see Fig.~\ref{fig:peakAKcomp}), the effects of an increase in dust opacity are likely to be negligible for our analysis. Moreover, \citet{Ascenso2013} show that the mid-infrared extinction law towards the FeSt 1-457 core and B59 are not significantly different to that derived for diffuse regions. This similarity also limits the variation in dust properties that is plausible for cores in the Pipe nebula.

Towards the south, the $A_K$ values in the HP2 map are lower than those in the extinction map. Since this is the direction towards the Galactic plane, the lines of sight are more likely to be contaminated with material associated with structures or clouds other than the Pipe nebula. The column densities in the southern areas of the map could be underestimated because several layers of dust are observed along the line of sight that differ in, for example, their temperature \citep[see, e.g.][]{Shetty2009b}, composition, or size distribution.

\begin{figure}
  \resizebox{\hsize}{!}{\includegraphics{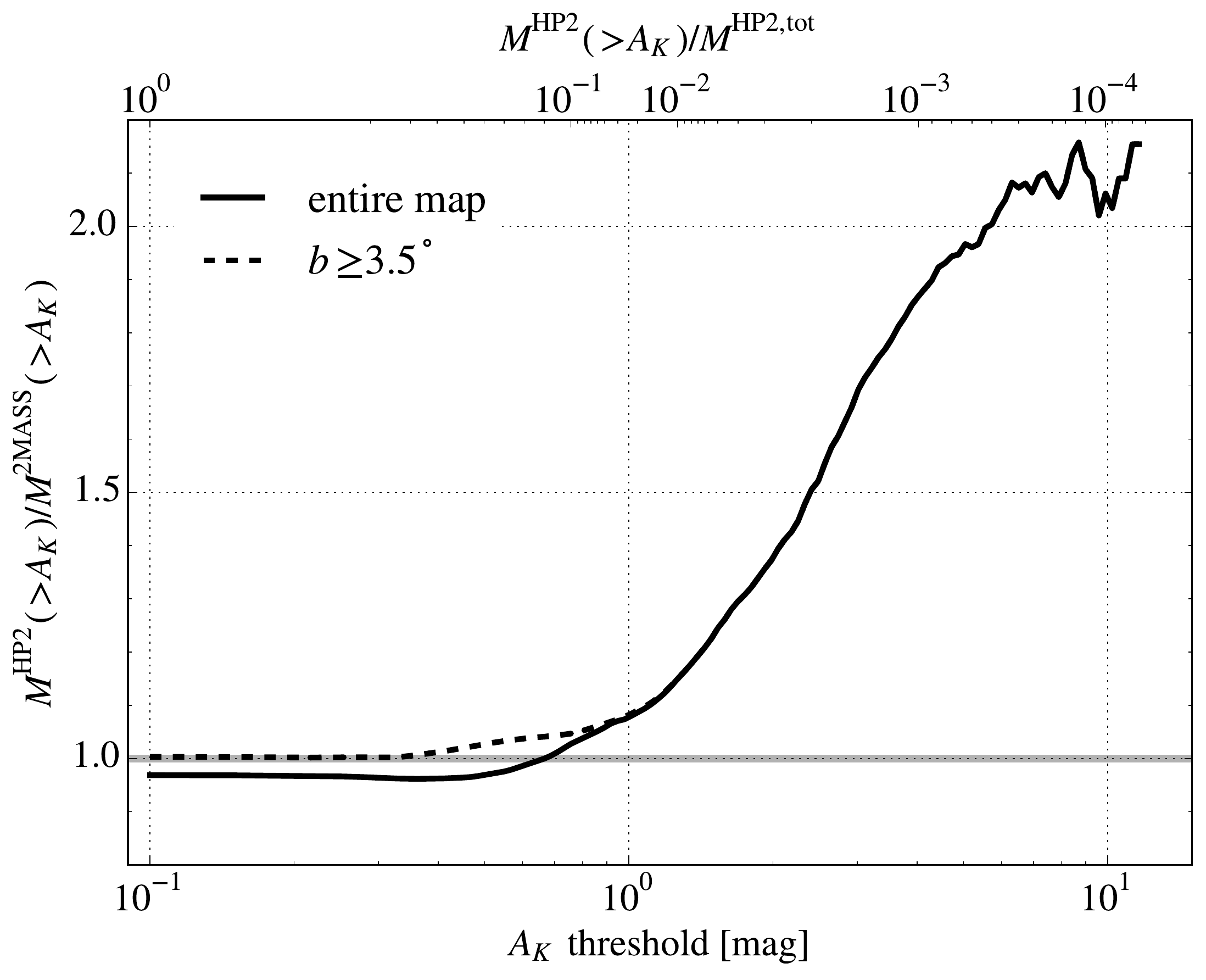}}
  \caption{Ratio between masses above an $A_K$ threshold derived from the HP2 and extinction map as a function of the threshold. The upper x~axis gives the fraction of the total mass in the map that is above the threshold.}
  \label{fig:Mratthres}
\end{figure}

The following analysis aims at quantifying the significance of the differences described in the previous paragraph and at evaluating whether a relevant fraction of the cloud mass exists in the form of substructure that is not traced by the extinction map. Since the extinction map is based on measurements towards point-like objects, it could be missing the small-scale structure of the cloud if individual pixels are not sampled with a sufficient number of background stars. We have calculated the mass contained in the area above a certain column-density threshold in the HP2 map. Extinction values were converted to surface mass density via the relation
\begin{equation}
\Sigma = \mu m_p \beta_K A_K
\end{equation}
where $\mu$ is the mean molecular weight, $m_p$ is the proton mass, and $\beta_K$ is the gas-to-dust ratio $N_H / A_K$. We assumed $\mu=1.37$ and $\beta_K= 1.87 \times 10^{22}\,\mathrm{cm}^{-2}\,\mathrm{mag}^{-1}$. The latter is based on canonical values for the diffuse ISM ($N_H / E(B-V) = 5.8 \times 10^{21}\,\mathrm{cm}^{-2}\,\mathrm{mag}^{-1}$, $R_V = A_V / E(B-V) = 3.1$, $A_K / A_V = 0.1$) as reported by \citet{Bohlin1978} and \citet{Cardelli1989}. Nevertheless, the gas-to-dust ratio $N_H / A_K$ is also valid in regions characterised by higher densities due to the insensitivity of the NIR extinction law to $R_V$.
The total mass of the region was calculated by integrating over the surface mass density within the region. We considered only pixels that show column densities above the threshold in the HP2 map, and compared the mass contained in these pixels for both the HP2 and the extinction map ($M^{\mathrm{HP2}} (>A_K)$ and $M^{\mathrm{2MASS}} (>A_K)$, respectively). The result is shown in Fig.~\ref{fig:Mratthres}. At the lowest threshold, the calculated masses correspond to the total mass in the map $M^{\mathrm{HP2, tot}}$. Here, the mass ratio is lower than one due to the lines of sight towards the south. If the southernmost ($b<3.5^{\circ}$) areas are excluded from the analysis (dashed line in Fig.~\ref{fig:Mratthres}), the mass ratio is close to one for low $A_K$ thresholds. As the threshold increases, the mass ratio increases continuously. At a threshold of $\sim 0.9$\,mag, the masses above the threshold differ by 5\% and correspond to $\sim$ 5\% of the total mass in the HP2 map. Similarly, at a threshold of $\sim 6$\,mag, we are probing $\sim$ 0.06\% of the total mass in the HP2 map and the masses differ by a factor of two. This suggests that although the extinction map underestimates column densities (and therefore masses) significantly in the highest column-density regions of the cloud, these regions represent only a very small fraction of the total mass of the cloud. We can thus also exclude a scenario in which the Pipe nebula contains substructure that is not traced by the extinction map and contributes substantially to the mass of the cloud.

\begin{figure}
  \resizebox{\hsize}{!}{\includegraphics{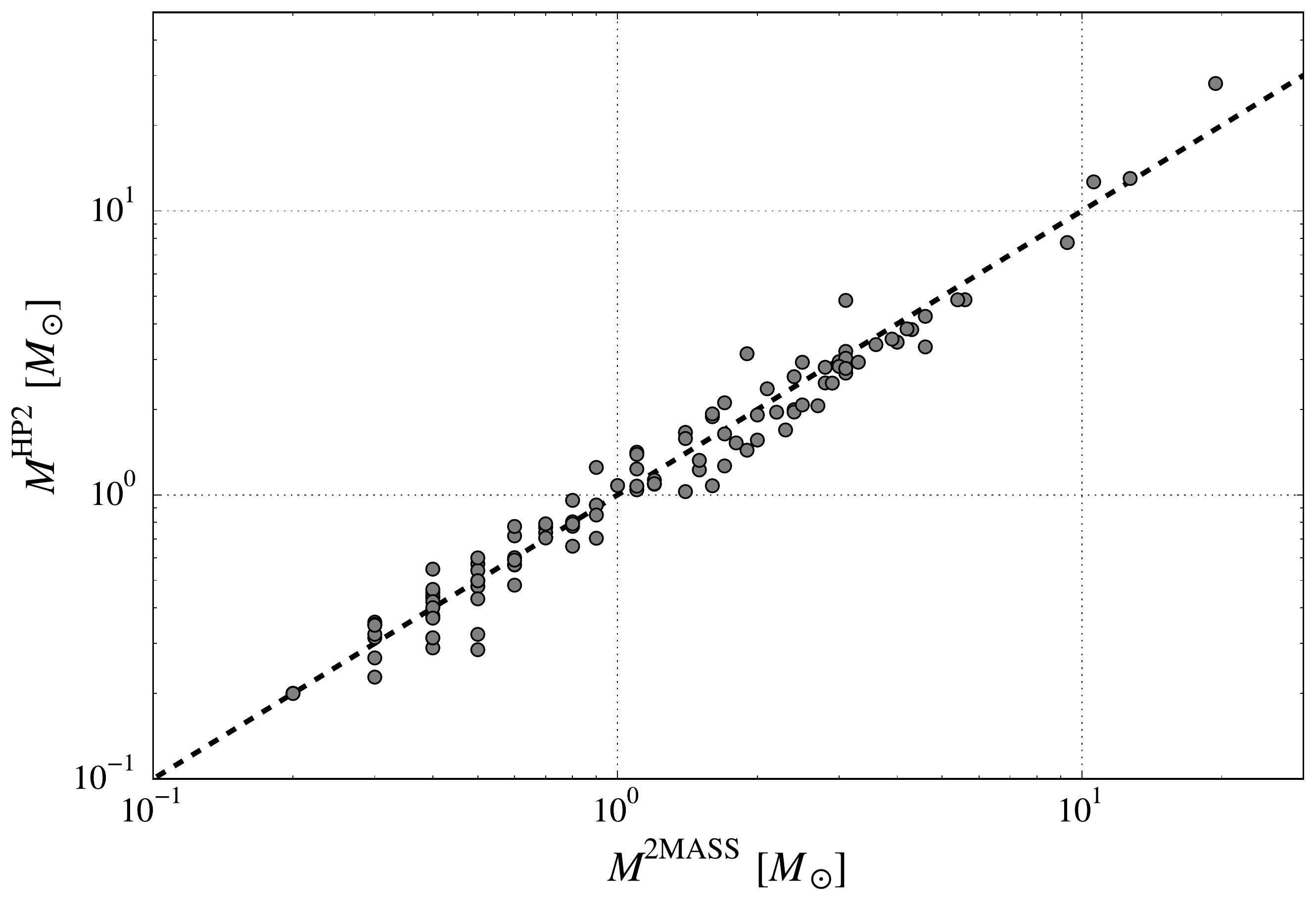}}
  \caption{Comparison of core masses derived by \cite{Rathborne2009} $M^{\mathrm{2MASS}}$ and from the HP2 map $M^{\mathrm{HP2}}$. The dashed line indicates equality.}
  \label{fig:Mcomp}
\end{figure}

Finally, we compared peak column densities and masses for the R09 core sample as derived from the HP2 and extinction maps. As shown in Fig.~\ref{fig:peakAKcomp}, the peak column densities agree well over a large range. Only the values at high $A_K$ deviate, as expected from our previous tests. The R09 core masses, however, are in excellent agreement over the entire mass range (see Fig.~\ref{fig:Mcomp}). Thus, the differences between the HP2 and extinction map appear to be largely irrelevant when integrated over core regions despite the oftentimes large deviations observed for individual pixels at high column densities.

\section{Discussion}

In addition to column density, dust-emission maps also provide information on the effective-temperature distribution of the mapped region, which is not possible with extinction measurements. How effective-dust-temperature measurements relate to physical properties of the observed medium is generally not clear, however. The complications associated with this relationship include the a priori unclear correspondence between dust and gas temperatures, the confusion of several independent structures along the same line of sight, and systematic effects introduced by the MBB model. We therefore aimed to assess the value of effective-dust-temperature maps in studies of nearby molecular clouds, in particular regarding the thermodynamics and boundaries of dense cores.

Observations of dust emission are capable of providing information on the thermodynamic state of a region only if dust and gas temperatures are correlated. We expect to observe this correlation in regions of high volume density, where gas and dust are thermally coupled \citep[e.g.][]{Goldsmith2001}. \cite{Forbrich2014} compare effective dust temperatures to kinetic gas temperatures $T_g$ derived from molecular line observations of NH$_3$ towards a sample of cores in the Pipe nebula. These authors find that the temperatures are correlated, with dust temperatures higher by ${\sim}$3\,K than the corresponding $T_g$ value. Similar results are reported for observations of dense gas in the NGC1333 region in Perseus \citep{Hacar2017}. Therefore, while effective dust temperatures are generally not equal to thermodynamic temperatures, they still allow for statements on the physical state of the observed matter in dense regions.

Previous observational and theoretical studies of cores inform our expectations of dust temperature variations in cores. Due to shielding from the interstellar radiation field (ISRF) provided by the outer layers of a core, the innermost parts can cool to lower temperatures. Overall, we thus expect an anti-correlation between column density and temperature in regions where shielding is effective. This phenomenon is predicted in models of dense cores \citep[e.g.][]{Evans2001, Zucconi2001, Galli2002, Bate2015} and low central core temperatures of ${\sim}10$\,K have been observed in numerous surveys \citep[e.g.][]{Myers1983, Crapsi2007, Roy2014, Lippok2016}. While an anti-correlation between $\tau_{850}$ and $T$ could also be caused by variations in dust properties, this effect is unlikely to affect the overall analysis carried out in this paper (see discussion in Sect.~\ref{subsec:compext}). In the following, we first searched for local minima in our HP2 temperature map and then analysed the relationship between column density and temperature inside and outside the R09 core regions to test the significance of dust-temperature information regarding the physical properties of cores.

\subsection{Dust temperature minima}

\begin{figure*}
  \resizebox{\hsize}{!}{\includegraphics{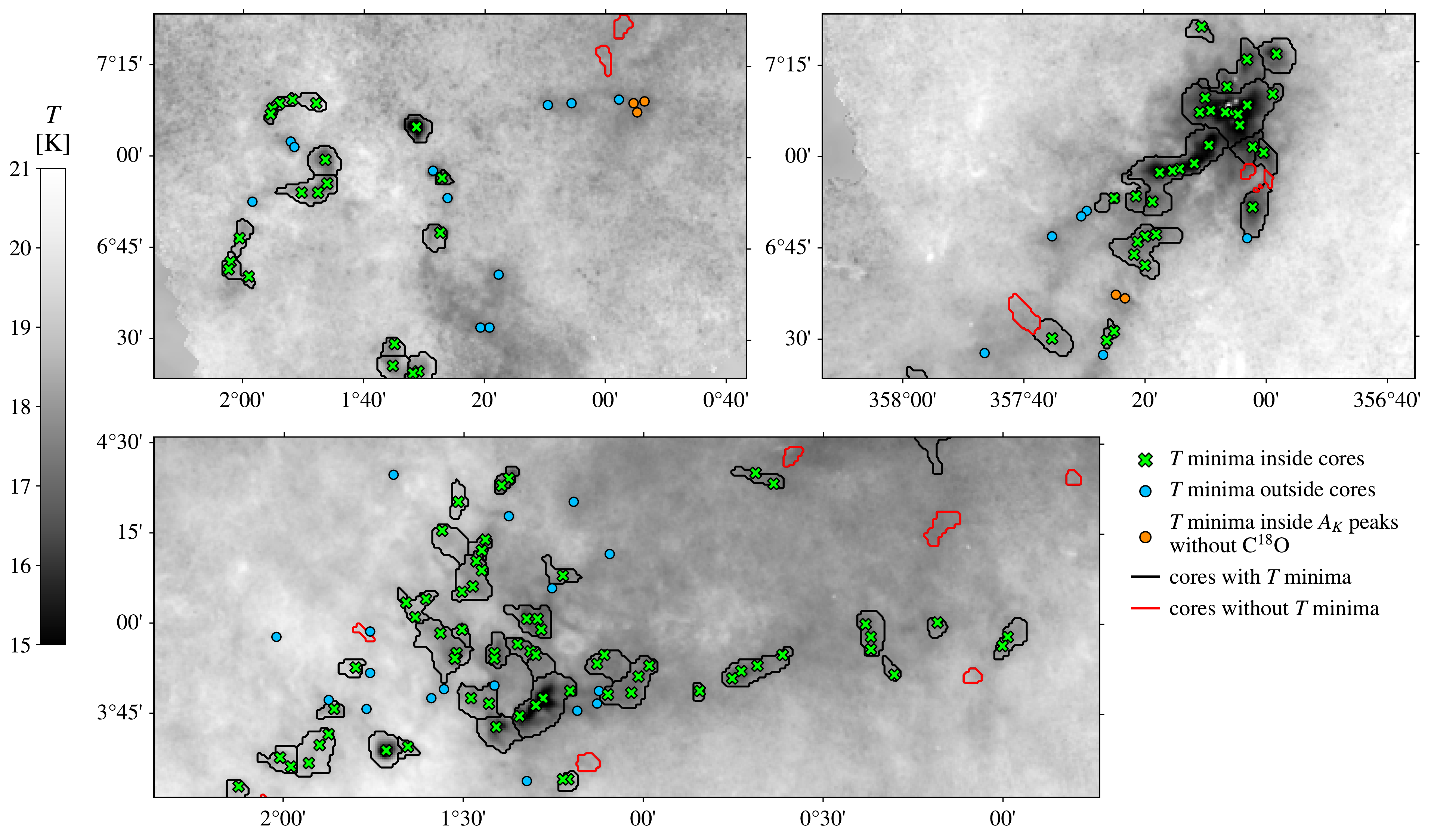}}
  \caption{Map of effective dust temperature $T$ for part of the Smoke region (\textit{top left}), part of the Stem region including B59 (\textit{top right}), and part of the Bowl region (\textit{bottom}), with $T$ minima and R09 cores indicated as described in the legend.}
  \label{fig:Tmin}
\end{figure*}

As a first analysis of the HP2 temperature map, we searched for local minima and correlated them with the R09 core sample. To this end, we first removed the large-scale background and reduced small scale noise in the inverted $T$ map. To achieve the former, we estimated the local background by applying a Gaussian filter with a standard deviation of ten~pixels to the map and subtracted it. To achieve the latter, we applied a Gaussian filter with a standard deviation of one~pixel. By dividing the resulting map by the temperature error map, we obtained the equivalent of a signal-to-noise (S/N) value for each pixel. The signal corresponds to the temperature difference to the local background, and the noise to the error given in the HP2 map. Next, we employed a simple method to identify local maxima in the S/N map: A maximum filter was applied to the map and any pixels with identical values in the original and filtered S/N map were considered potential local maxima. This search was restricted to the areas covered by \textit{Herschel} due to the higher resolution we achieve there. Finally, our sample of local maxima consisted only of those potential maxima with an S/N value above a particular threshold. We chose the maximum-filter kernel size and the S/N threshold by visual inspection of the $T$ minima found. In order to identify cold structures as individual minima also in crowded regions such as the Bowl of the Pipe nebula, we set the kernel size to two~pixels. This relatively small filter kernel necessitates a relatively high S/N threshold of ten to exclude the large number of potential maxima found in noisy regions, for example the edges of \textit{Herschel} fields.

The spatial distribution of $T$ minima and R09 cores is shown in three regions of the Pipe nebula in Fig.~\ref{fig:Tmin}. The population of $T$ minima appears to follow the filamentary structure of the cloud that is visible in the column-density and temperature maps. Two obvious exceptions are located in a noisy area of the S/N map in the north of the central \textit{Herschel} field (outside the areas shown in Fig.~\ref{fig:Tmin}). For the comparison with our previously defined core sample, only the 109~R09 cores that lie within the area covered by \textit{Herschel} were considered. Since this comparison is associated with several caveats (see below), we restrict the following discussion to largely qualitative arguments. We found that the majority of R09 cores (93 of 109 cores, $\sim$ 85\%) contains at least one $T$ minimum. Furthermore, a considerable amount of minima is located outside the R09 core regions (69 of 248 $T$ minima, $\sim$ 28\%). Generally, our results are therefore compatible with the literature: most R09 cores exhibit a drop in temperature with respect to their environment, which is indicative of an anti-correlation between column density and temperature.

Our analysis of $T$ minima is associated with several complications, however, which we outline briefly. The sample of R09 cores and $T$ minima is based on two different maps and two different extraction methods. Due to the higher resolution of the HP2 map, we could detect minima that are not resolved as dense structures in the extinction map. Moreover, a temperature gradient along the line of sight could lead to an overestimate of $T$, while several relatively diffuse structures along the line of sight could appear as a dense core in the extinction map. \citet{Rathborne2009} additionally employed molecular line data to alleviate the latter issue. In principle, the observation of both C$^{18}$O and a $T$ minimum can be interpreted as an indication of high volume densities. Interestingly, we found five $T$ minima in regions that were excluded from the R09 core sample because no C$^{18}$O was detected towards these areas. This apparent contradiction originates from the application of two different sets of criteria to find dense structures, which itself is a result of the lack of a physically motivated core definition. Similarly, the population of $T$ minima not associated with R09 cores can be interpreted as calling for a core definition based on physical properties of the region. The fact that we found a significant overlap between our sample of $T$ minima and R09 cores, despite these caveats, suggests that the HP2 dust temperature map indeed contains information on the physical state of the observed medium.

\subsection{$\tau_{850}$-$T$ relation}

\begin{figure}
  \resizebox{\hsize}{!}{\includegraphics{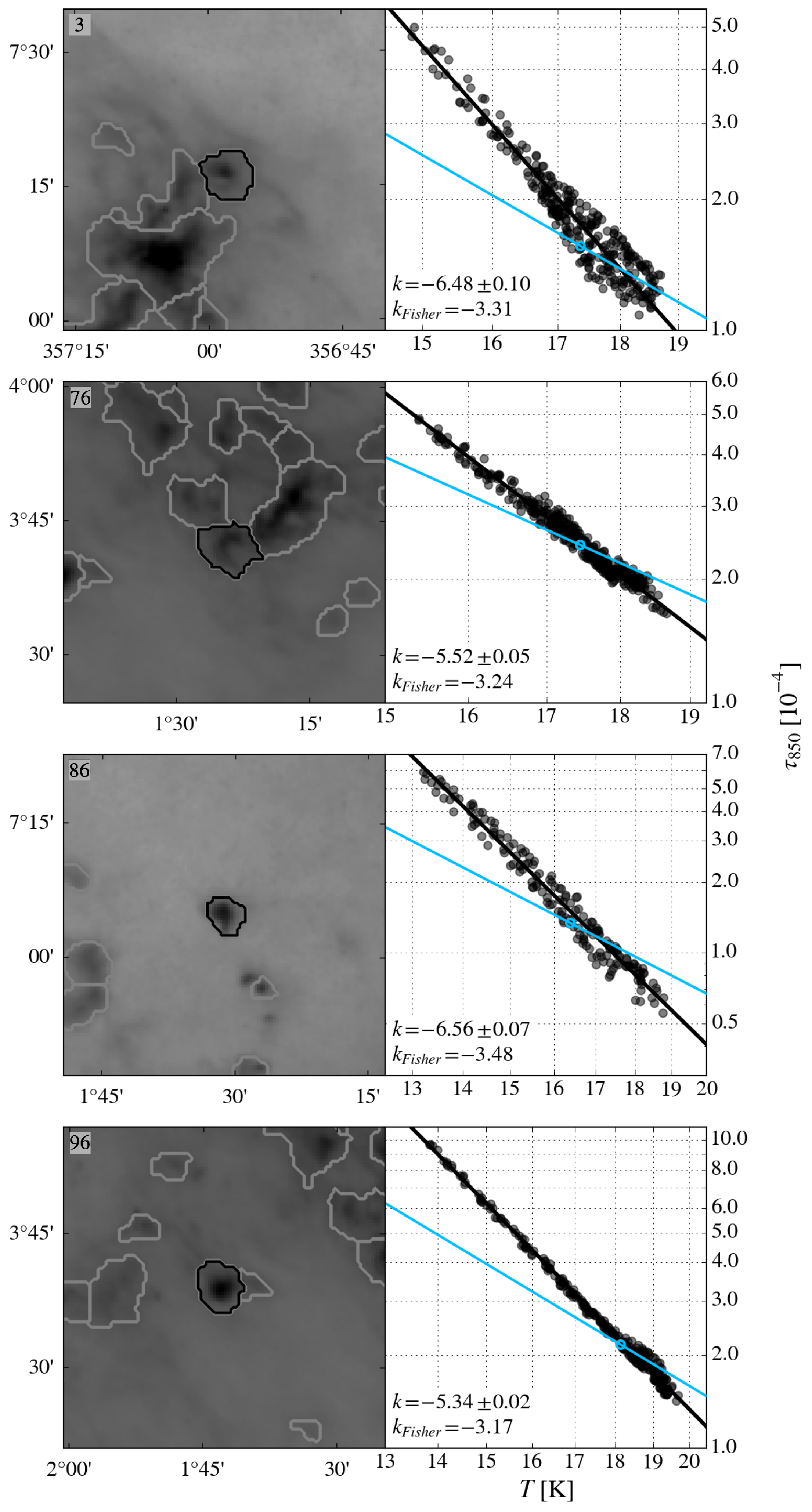}}
  \caption{Column density $\tau_{850}$ as a function of effective temperature $T$ for four R09 cores with clear linear relations. The left panels show the column-density map with the core indicated by black contours. The R09 core number is given in the top left corner. Other R09 cores in the field are indicated by grey contours. In the right panels, each data point corresponds to one pixel in the HP2 maps. The black line is a power-law fit to the data points. The blue line is a power-law function with the exponent $k_{Fisher}$ and the normalisation chosen such that the function passes through the reference data point indicated by a blue circle.}
  \label{fig:crFisher}
\end{figure}

\begin{figure}
  \resizebox{\hsize}{!}{\includegraphics{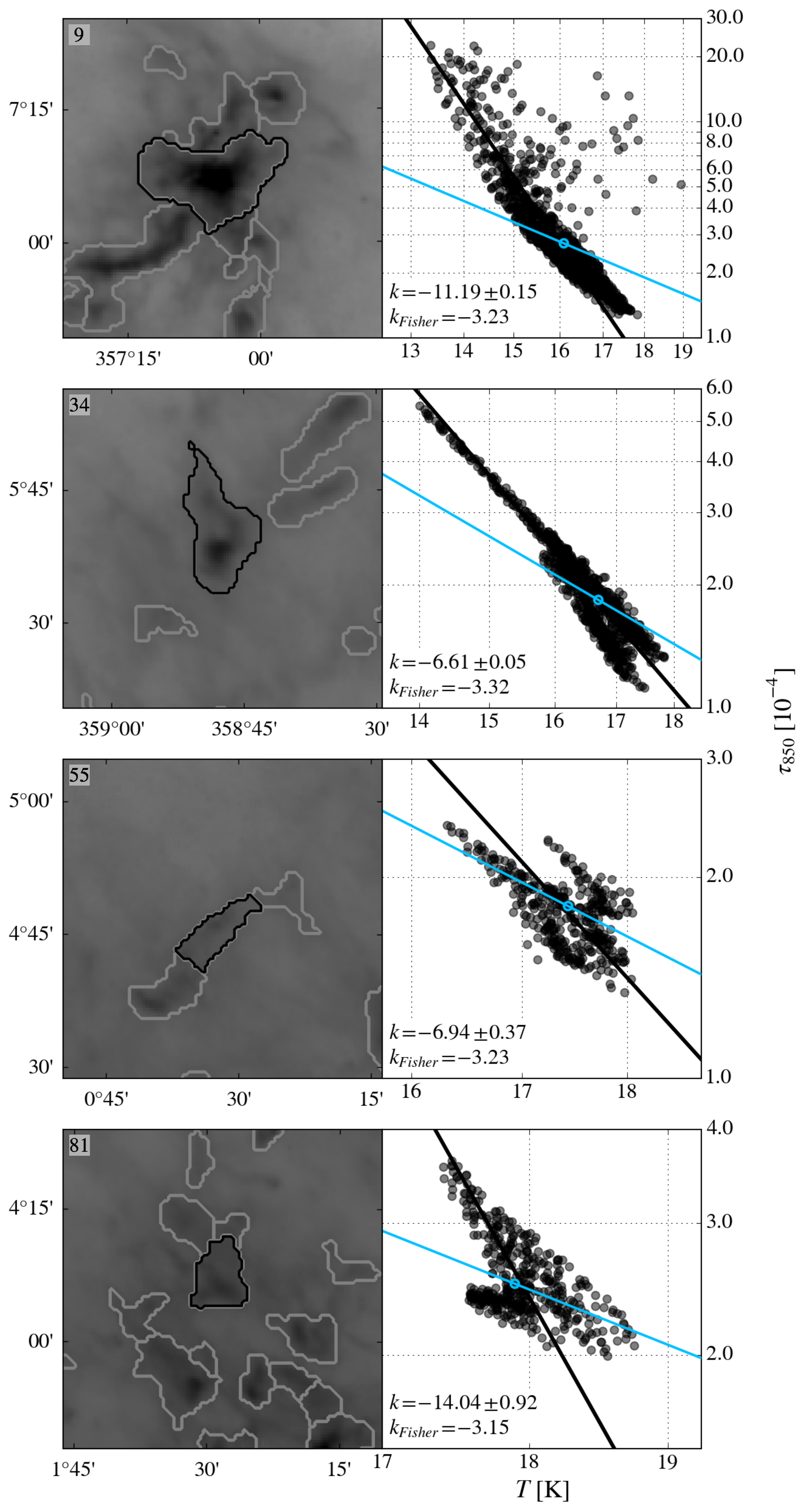}}
  \caption{Column density $\tau_{850}$ as a function of effective temperature $T$ for four R09 cores with complex relations. The left panels show the column-density map with the core indicated by black contours. The R09 core number is given in the top left corner. Other R09 cores in the field are indicated by grey contours. The right panels follow the same colour coding as Fig.~\ref{fig:crFisher}.}
  \label{fig:crdoubleFisher}
\end{figure}

The HP2 maps allow us to quantify and investigate in more detail the column-density-temperature anti-correlation indicated by our previous results. As a first step, the relation between $\tau_{850}$ and $T$ was examined within each R09 core. Again, only R09 cores located within the \textit{Herschel} coverage were considered. For these R09 cores, the majority exhibit a clear anti-correlation between $\tau_{850}$ and $T$ (see Fig.~\ref{fig:crFisher}). A number of R09 cores show more complex behaviour (see Fig.~\ref{fig:crdoubleFisher}), including some where two distinct relations appear to overlap within one R09 core. In B59 (first panel in Fig.~\ref{fig:crdoubleFisher}), several data points are located off the main relation towards higher temperatures, which is a consequence of the heating induced by the YSOs inside the core region. Overall, this analysis thus supports our previous conclusions: We observe an anti-correlation between column density and temperature in dense regions due to the higher degree of shielding from the ISRF.

\begin{figure}
  \resizebox{\hsize}{!}{\includegraphics{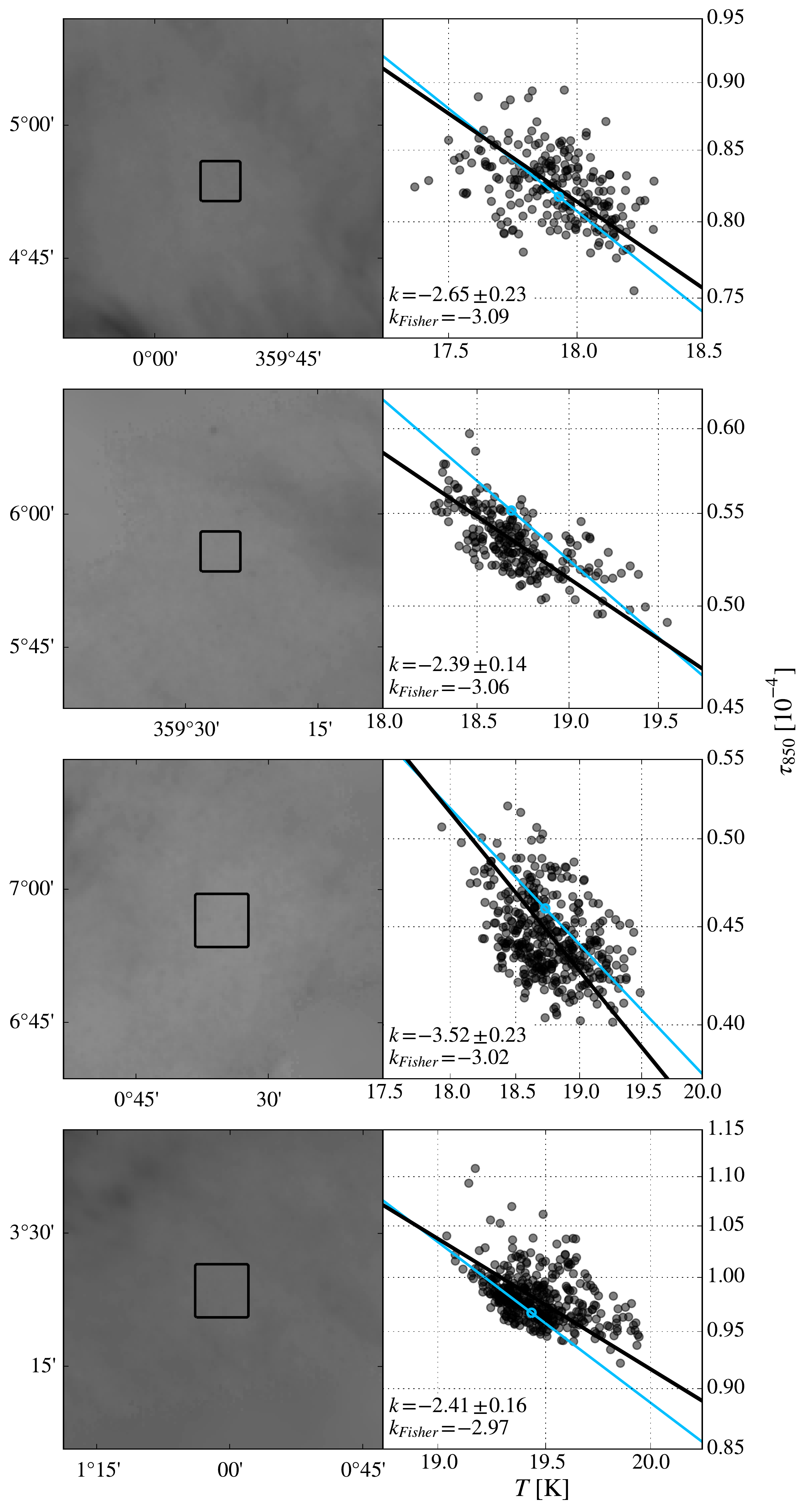}}
  \caption{Column density $\tau_{850}$ as a function of effective temperature $T$ for all four test areas. The left panels show the column-density map with the test areas indicated by black contours. The right panels follow the same colour coding as Fig.~\ref{fig:crFisher}.}
  \label{fig:taFisher}
\end{figure}

\begin{figure}
  \resizebox{\hsize}{!}{\includegraphics{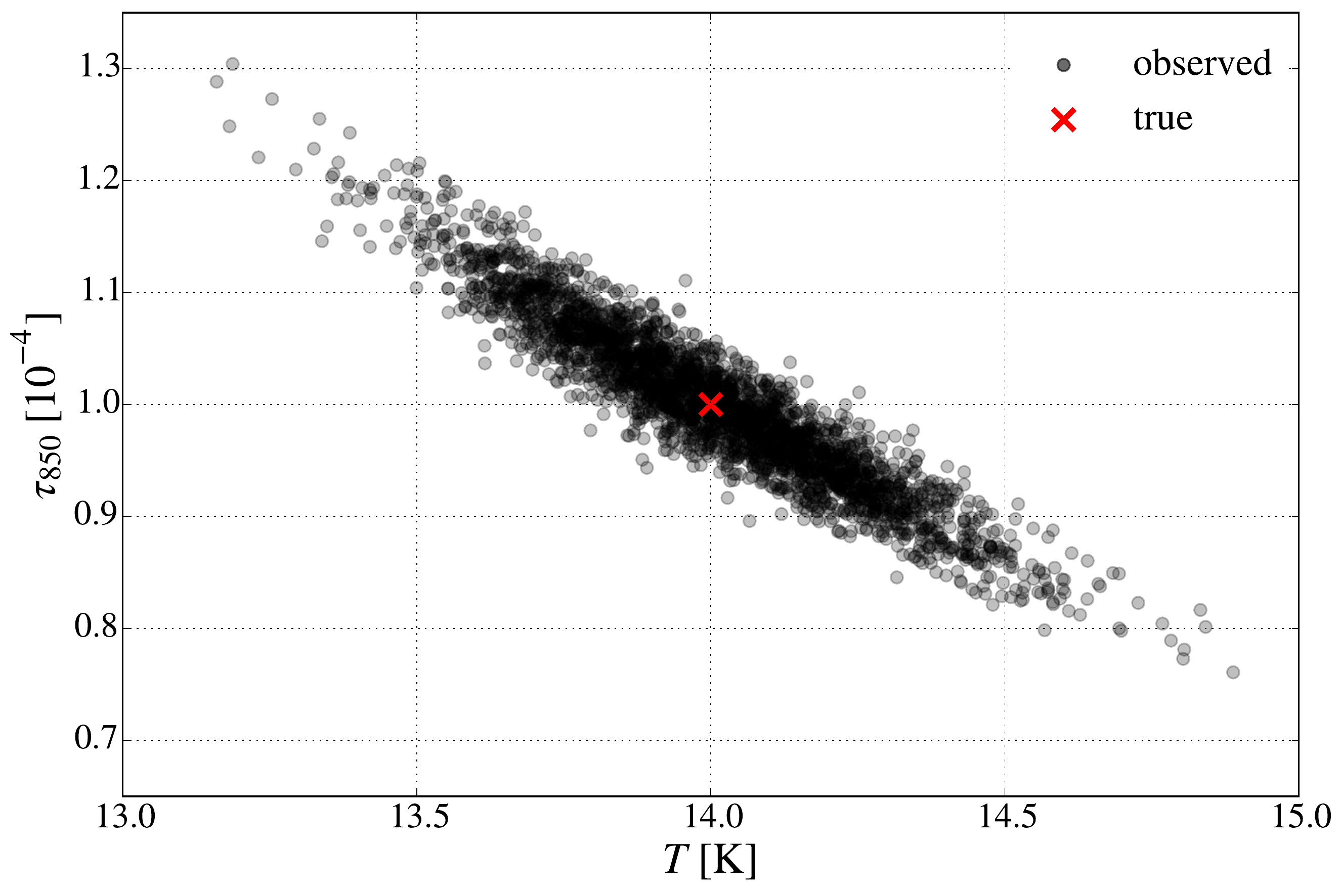}}
  \caption{Observed optical depth $\tau_{850}$ and effective temperature $T$ for the same set of true parameters (indicated by a red X) and 3000 different realisations of flux values assuming a relative flux error of 5\,\%. The spectral index $\beta$ is set to 1.6.}
  \label{fig:fitdeg_ex}
\end{figure}

In order to test the significance of this finding, we also inspected four square-shaped test areas within the \textit{Herschel} coverage. These areas were selected to appear uniform in both the~$\tau_{850}$ and~$T$ map, which implies that the regions are devoid of dense structure and should therefore not show indications of shielding. However, we found an anti-correlation between $\tau_{850}$ and $T$ in all four test areas (see Fig.~\ref{fig:taFisher}). This result is in clear contradiction to the expectations we framed based on the literature and our previous analyses if we assume that the anti-correlation is caused by the effect of shielding from the ISRF. A negative correlation between $\tau_{850}$ and $T$ can, however, also be the result of a degeneracy between the two parameters in the MBB fit (see, e.g. \citealp{Planck2011dust, Planck2014}). We illustrate this effect by assuming a set of true parameters to construct a spectral energy distribution, from which the true flux values in the PACS and SPIRE passbands can be deduced. The observed flux values were then sampled from Gaussian distributions with the true values at their centres and an assumed variance. Using the same model and fitting algorithm as described in Sect.~\ref{sec:methods}, the observed $\tau_{850}$ and $T$ values were derived. An example with 3000 realisations of the same set of true parameters is shown in Fig.~\ref{fig:fitdeg_ex}. The distribution of observed values has a different shape depending on the choice of true parameters and errors.

Consequently, the next task was to find a metric that is capable of distinguishing an anti-correlation caused by physical processes from an anti-correlation caused by the systematic MBB-fit degeneracy. For most R09 cores, the relation between $\tau_{850}$ and $T$ is well-described by a linear function in a log-log plot. We therefore fitted a power-law function to the data using an orthogonal distance regression algorithm taking the errors in both $\tau_{850}$ and $T$ into account (black lines in right panels of Figs.~\ref{fig:crFisher}, \ref{fig:crdoubleFisher}, and~\ref{fig:taFisher}). The distribution of the best-fit exponents $k$ shows that the test areas have shallower relations than the majority of cores. In order to compare these exponents with anti-correlations caused entirely by the degeneracy in the MBB fit, we employed the Fisher information matrix \citep{Fisher1922}. The Fisher information matrix measures the ability of a data set to constrain the parameters of a model used to describe the data set. The Cram\'{e}r-Rao bound \citep{Rao1945, Cramer1946} states that the inverse of the Fisher information matrix is a lower limit to the covariance matrix associated with a set of parameters. Here, this allows us to derive an estimate of the slope that can be induced by the MBB fit, $k_{Fisher}$\footnote{The Fisher slope, by definition, is the shallowest slope that can be induced by the MBB fit. However, it is still a useful limiting value in this context since $k_{Fisher}$ is close to the slope caused by the MBB-fit degeneracy if the model describes the data well.} (blue lines in right panels of Figs.~\ref{fig:crFisher}, \ref{fig:crdoubleFisher}, and~\ref{fig:taFisher}).

For each R09 core and test area, we calculated the Fisher information matrix at one reference pixel which we chose to be the pixel at the median $T$ in that core or test region. We assumed the likelihood $\mathcal{L} \propto \exp^{-\chi^2/2}$ and calculated the Hessian matrix of the log-likelihood with respect to the two free fit parameters $\tau_{850}$ and $T$. The Fisher matrix was obtained by averaging the Hessian matrix elements over 100 data realisations drawn from a Gaussian distribution centred on the observed flux and with a standard deviation corresponding to the flux error. From the inverse of the Fisher matrix, we can derive an error ellipse whose inclination angle $\alpha$ we compared to the exponents $k$. Since $\alpha$ is defined on linear scales in $\tau_{850}$ and $T$, while $k$ represents the slope in a log-log plot, we transformed the inclination angle by deriving the corresponding slope in a log-log plot at the reference pixel with the effective temperature $T_0$ and column density $\tau_0$ via
\begin{equation}
k_{Fisher} = \frac{\mathrm{d}(\log(\tau))}{\mathrm{d}(\log(T))}\  \Big|_{(\tau_0, T_0)} = \left( \frac{\tau}{T} \frac{\mathrm{d}\tau}{\mathrm{d}T} \right)\  \Big|_{(\tau_0, T_0)} = \frac{\tau_0}{T_0} \tan(\alpha).
\end{equation}
If a core or test area exhibits a slope $k$ that is shallower than or compatible with $k_{Fisher}$, the observed anti-correlation can be explained entirely by the degeneracy effect of the MBB fit. All regions with slopes shallower than or within 3$\sigma$ of $k_{Fisher}$ were therefore labelled as candidate fake cores.

\begin{figure}
  \resizebox{\hsize}{!}{\includegraphics{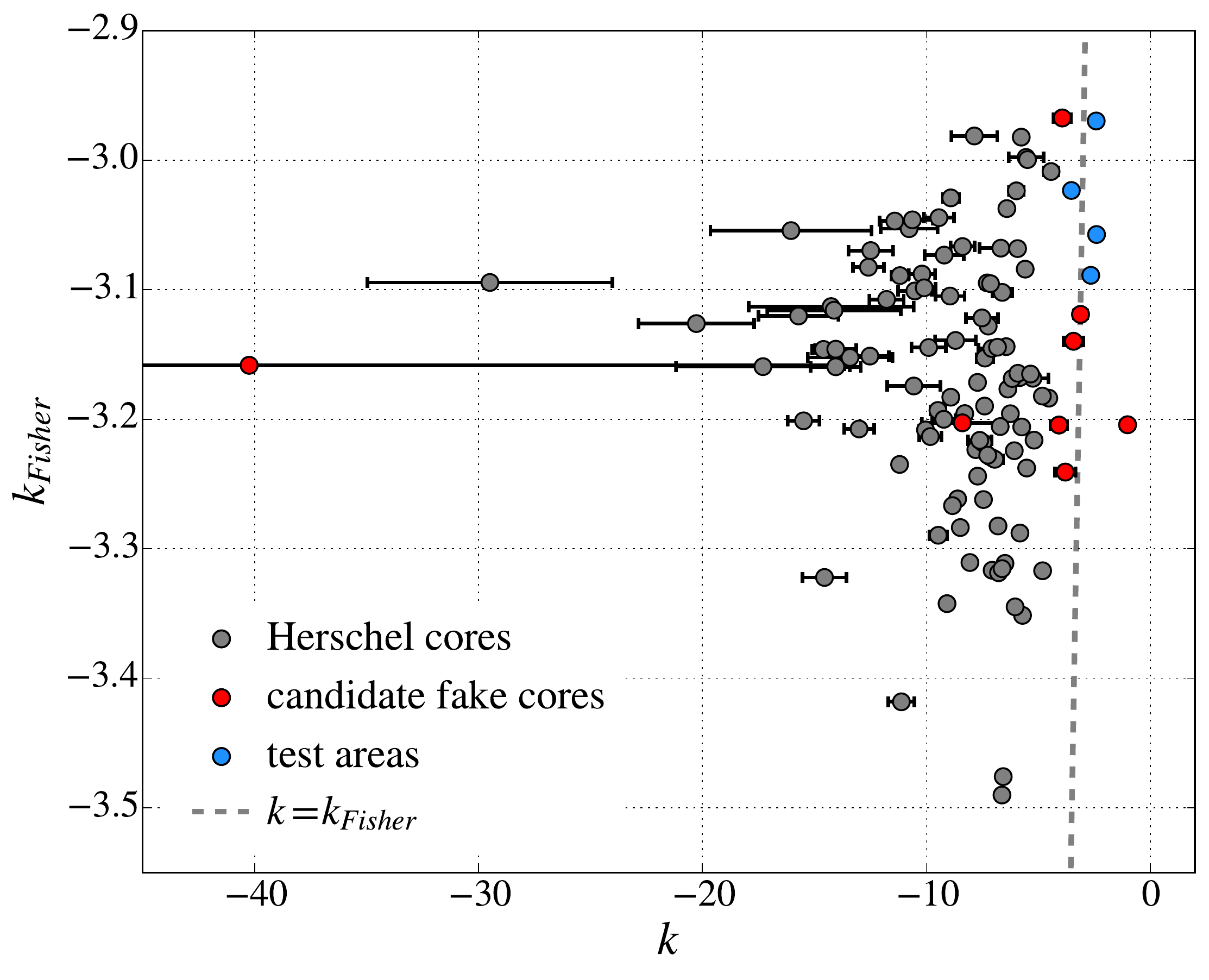}}
  \caption{Comparison of slopes derived from a fit to the column-density and effective-temperature data within R09 cores $k$ and derived from the Fisher information matrix $k_{Fisher}$. Grey, red, and blue circles correspond to all R09 cores within the \textit{Herschel} coverage, the candidate fake cores as a subset of these, and the test areas, respectively. The dashed line indicates equality.}
  \label{fig:kFisher}
\end{figure}

We identified eight of 109 R09 cores in the \textit{Herschel} coverage to be candidate fake cores (see Fig.~\ref{fig:kFisher}). In our sample of test areas, all four regions were labelled as candidate fake cores. The method is thus capable of recognising the lack of dense material in the test areas, all of which have comparably shallow slopes. At the same time, ${\sim}$93\% of R09 cores were correctly identified as containing dense structure. These results confirm our earlier findings and show that the thermodynamics of the medium are dominated by heating by the ISRF and shielding by the surrounding material in the majority of cores. The $\tau_{850}$-$T$ relation can be used consistently across the cloud to distinguish between core and diffuse regions in a physically motivated fashion.

The vast majority of R09 cores in the Pipe nebula are not associated with YSOs, and the conclusion that the cores are predominantly heated by the ISRF might therefore not be surprising. However, the B59 region hosts YSOs and while they cause higher dust temperatures in their vicinity, they do not alter the overall shape of the $\tau_{850}$-$T$ relation significantly. Thus, the dominant heating mechanism is also a question of spatial scale: on the scale of entire cores, the ISRF might be the dominant even in cores hosting YSOs that heat the ISM on smaller scales. However, B59 is the most massive core in the R09 sample and heating by YSOs could be more important in lower-mass cores.

\subsection{R09 core boundaries}

\begin{figure}
  \resizebox{\hsize}{!}{\includegraphics{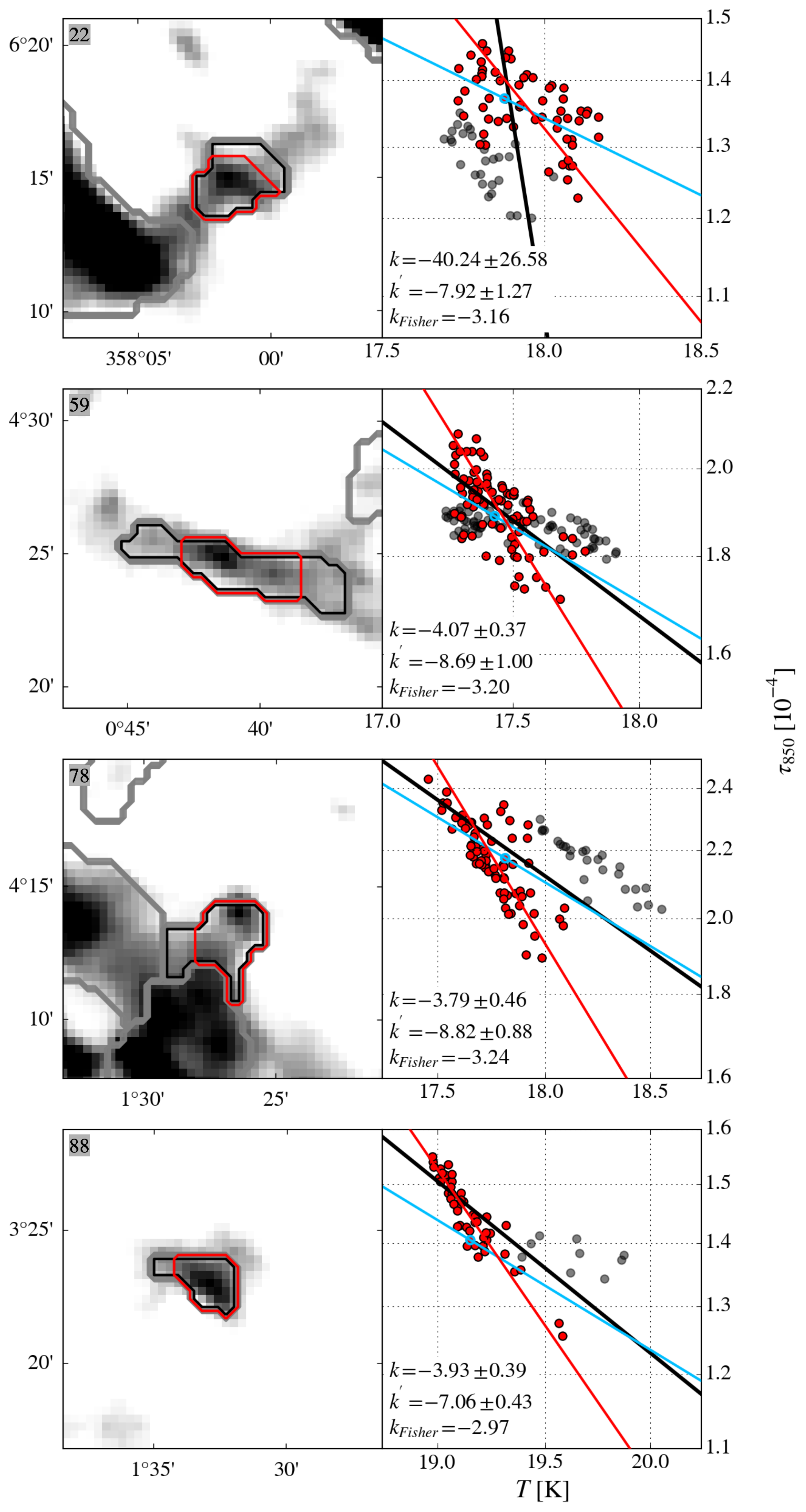}}
  \caption{Column density $\tau_{850}$ as a function of effective temperature $T$ for four candidate fakes cores with substructure. The left panels show the column-density map with the R09 core indicated by black and the adapted core boundaries by red contours. The R09 core number is given in the top left corner. Other R09 cores in the field are indicated by grey contours. To highlight substructures within the cores, the colour scale's upper and lower limits are set to the minimum and maximum $\tau_{850}$ value in the core. The right panels follow the same colour coding as Fig.~\ref{fig:crFisher}. Data points within the adapted boundaries are shown as red circles, and the corresponding power-law fit with the slope $k^{\prime}$ as a red line.}
  \label{fig:cfakesFisher}
\end{figure}

A closer look at the eight candidate fake cores revealed that in some cases the R09 core regions contain more than one structure and areas with comparably low column density. This is a result of the lower resolution of the extinction map which was used to define the R09 core boundaries. As a consequence, we expect these R09 cores to show steeper slopes if the core boundaries are adjusted appropriately. Indeed, in four of eight cases the boundaries can be adapted based on visual inspection to exclude diffuse regions or additional structures so that the resulting slope is too steep for the core to be considered a candidate fake core (see Fig.~\ref{fig:cfakesFisher}). As an example, we have considered R09 core~59. The HP2 column-density map shows substructure within the R09 core region, in particular structures with lower column densities towards the eastern and western edges of the region. In the comparison between $\tau_{850}$ and $T$ for this R09 core, two overlapping clouds of data points with different slopes are apparent. We found that removing the pixels towards the eastern and western edges of the region eliminates a significant part of the data points with the shallower $\tau_{850}$-$T$ relation. This suggests that the excluded pixels contain matter that is more diffuse than the central part of the R09 core.

The $\tau_{850}$-$T$ relation is evidently sensitive to region boundaries and our analysis thus suggests that it could be employed to refine the boundaries of dense structures. Setting core boundaries based on the anti-correlation between $\tau_{850}$-$T$ is one potential realisation of a physically motivated core definition that, however, requires further investigation. Developing a method using spatial, column-density, and temperature information in concert to derive core boundaries based on the physical properties of a region will be the focus of a forthcoming publication.

Four cores of the R09 sample (cores 38, 42, 50, and 56) remained classified as candidate fake cores in our analysis. Correspondingly, these regions fulfill the criteria to be considered dense based on C$^{18}$O observations and their morphology in an extinction map, while they do not fulfill our criteria based on dust emission measurements. Both sets of criteria are associated with their respective caveats. However, due to the location of the Pipe nebula in the sky, observations can be affected significantly by contributions from fore- and background material. In the case of dust emission measurements, the typically warmer medium in the fore- and background can cause an overestimate of effective dust temperatures \citep[see, e.g.][]{Shetty2009b}. We can therefore not exclude that the sample of candidate fake cores exhibits steep $\tau_{850}$-$T$ relations if only the core material is considered, since contributions from unrelated matter along the line of sight could influence our observations. Generally, we expect this effect to be less relevant for analyses of nearby molecular clouds located further away from the Galactic plane.

\section{Conclusions}

We present dust emission maps of the Pipe nebula region based on \textit{Herschel} and \textit{Planck} data with high resolution and large dynamic range in both optical depth and temperature. The maps are made publicly available at the CDS. A comparison with observations of NIR extinction in the same region shows that the emission and extinction maps are consistent. The emission map underestimates column densities towards the south of the region possibly due to line-of-sight confusion with warmer dust unrelated to the cloud, whereas the extinction map underestimates column densities towards the densest structures in the Pipe nebula due to a lack of background stars observed in the NIR. Nevertheless, masses derived from the two maps typically differ substantially only in areas of high column density, which do not contribute significantly to the total mass in the map. We also studied a sample of cores that was defined previously based on an extinction map of the region and found the core masses to be consistent between the HP2 and the extinction map.

We evaluated the relevance of the dust-temperature information in our HP2 maps by considering the temperature structure inside the core regions. The majority of cores is associated with at least one minimum in the temperature map, in agreement with theoretical predictions and observational studies reported in the literature. Within the cores and also in diffuse regions, column density and temperature show a negative correlation. We found that the slope of the anti-correlation is a good metric to distinguish between its two possible origins: in diffuse areas, the slope is comparably shallow and is caused by a systematic effect induced by the MBB fit. In core regions, the anti-correlation is steeper and the result of the physical processes that dominate the thermodynamics of dense structures in the Pipe nebula, namely heating by the ISRF and shielding by cloud and core material. Furthermore, our analysis indicates that core boundaries can be refined based on the column-density-temperature relation. We have thus taken a first step towards a physically motivated definition of dense structures that uses column-density, temperature, and spatial information in concert. Dust temperature maps clearly contain valuable information about the physical state of the observed medium and allow for deeper insights into the properties of molecular clouds and of the cores they host.

\begin{acknowledgements}
We thank the anonymous referee whose comments helped improve the manuscript. {\it Herschel} is an ESA space observatory with science instruments provided by European-led Principal Investigator consortia and with important participation from NASA. This publication makes use of data products from the Two Micron All Sky Survey, which is a joint project of the University of Massachusetts and the Infrared Processing and Analysis Center/California Institute of Technology, funded by the National Aeronautics and Space Administration and the National Science Foundation. This work is part of the research programme VENI with project number 639.041.644, which is (partly) financed by the Netherlands Organisation for Scientific Research (NWO). AH thanks the Spanish MINECO for support under grant AYA2016-79006-P. This research made use of Astropy, a community-developed core Python package for Astronomy \citep{Astropy2013}, NumPy \citep{NumPy2011}, and Matplotlib \citep{Hunter2007}.
\end{acknowledgements}

\bibliographystyle{aa}
\bibliography{mybib}

\begin{appendix}

\section{Additional maps in multi-layer figures}
\label{sec:applayers}

\begin{figure}[!h]
  \resizebox{\hsize}{!}{\includegraphics{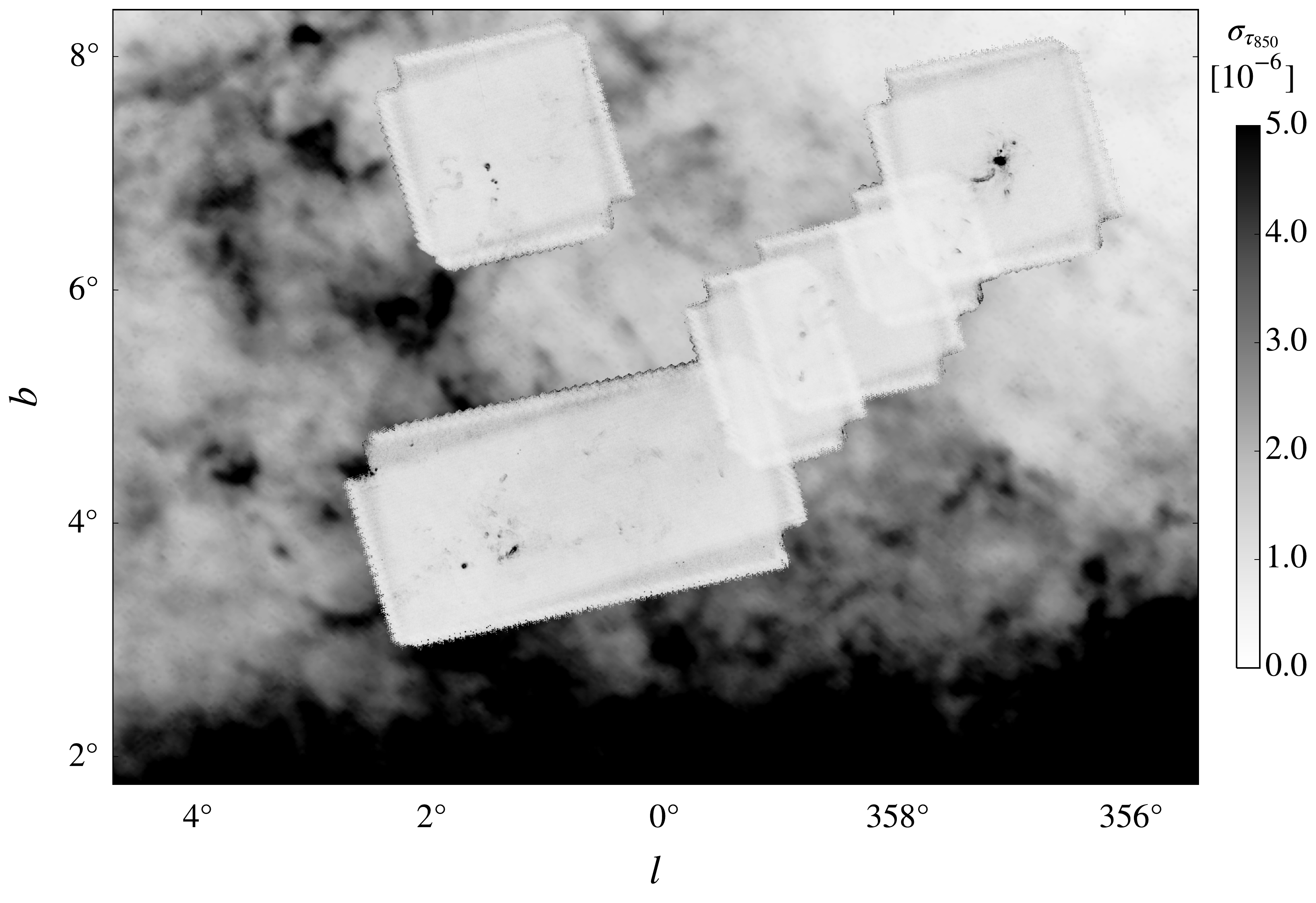}}
  \caption{Maps of the optical-depth error $\sigma_{\tau_{850}}$ derived from dust emission.}
  \label{fig:tauerrmap}
\end{figure}

\begin{figure}[!h]
  \resizebox{\hsize}{!}{\includegraphics{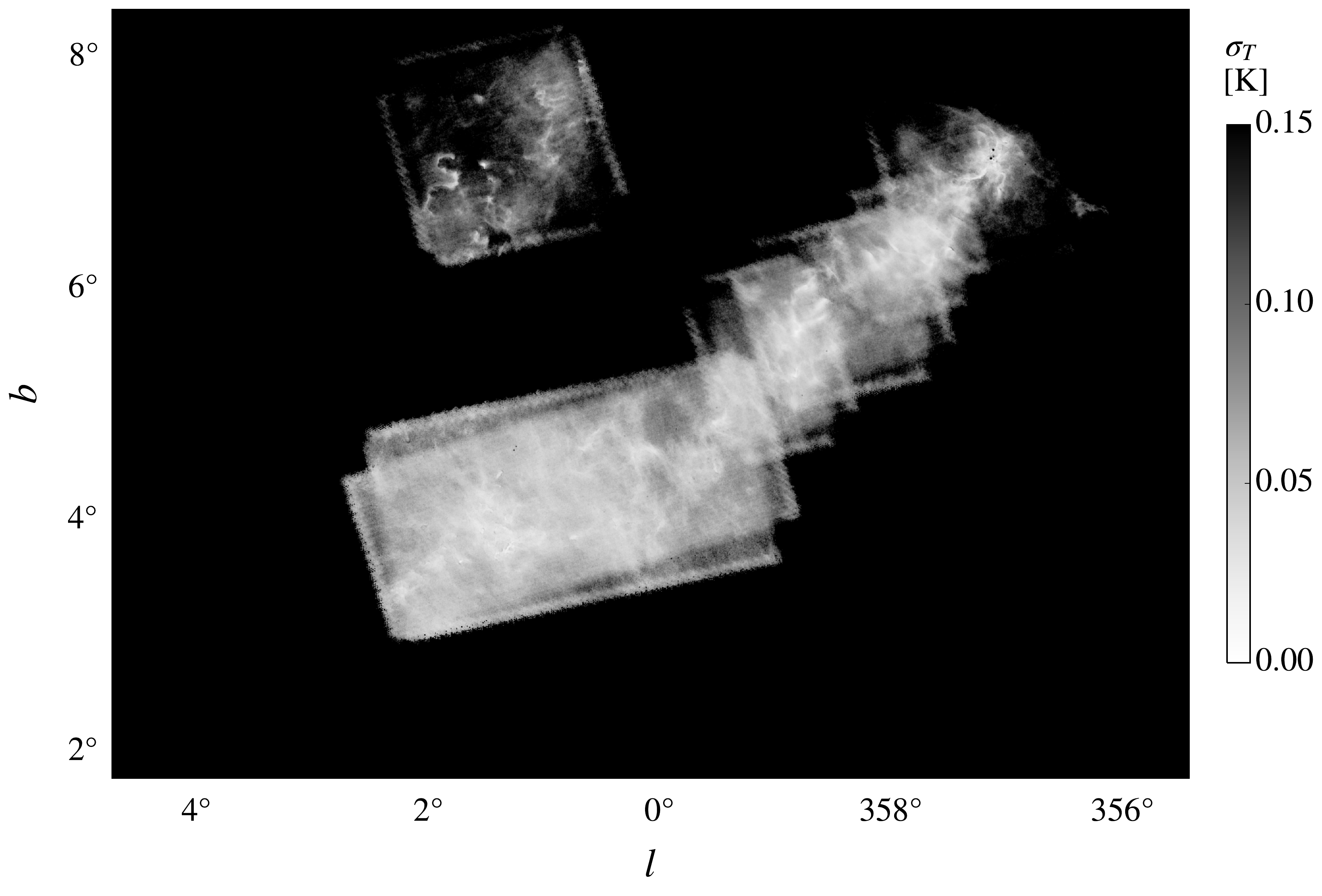}}
  \caption{Maps of the effective-temperature error $\sigma_T$ derived from dust emission.}
  \label{fig:Terrmap}
\end{figure}
\vfill\null
\break

\section{Additional distributions for global statistical analysis}
\label{sec:appdist}

\begin{figure}[!h]
  \resizebox{\hsize}{!}{\includegraphics{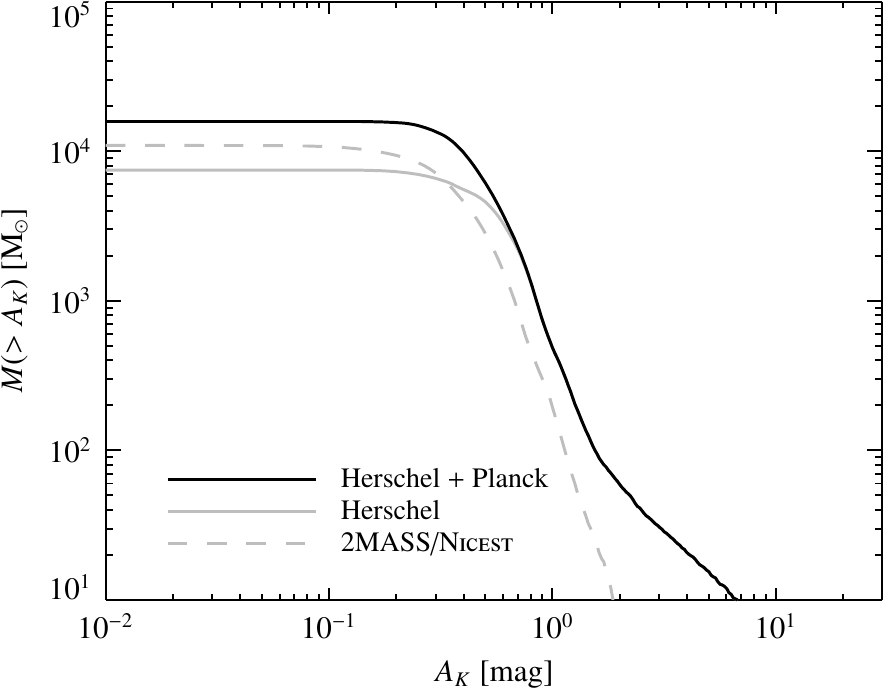}}
  \caption{Integrated mass $M(>A_K)$ as a function of extinction threshold $A_K$. The colour coding of the lines is as in Fig.~\ref{fig:Sdist}.}
  \label{fig:Mdist}
\end{figure}

\begin{figure}[!h]
  \resizebox{\hsize}{!}{\includegraphics{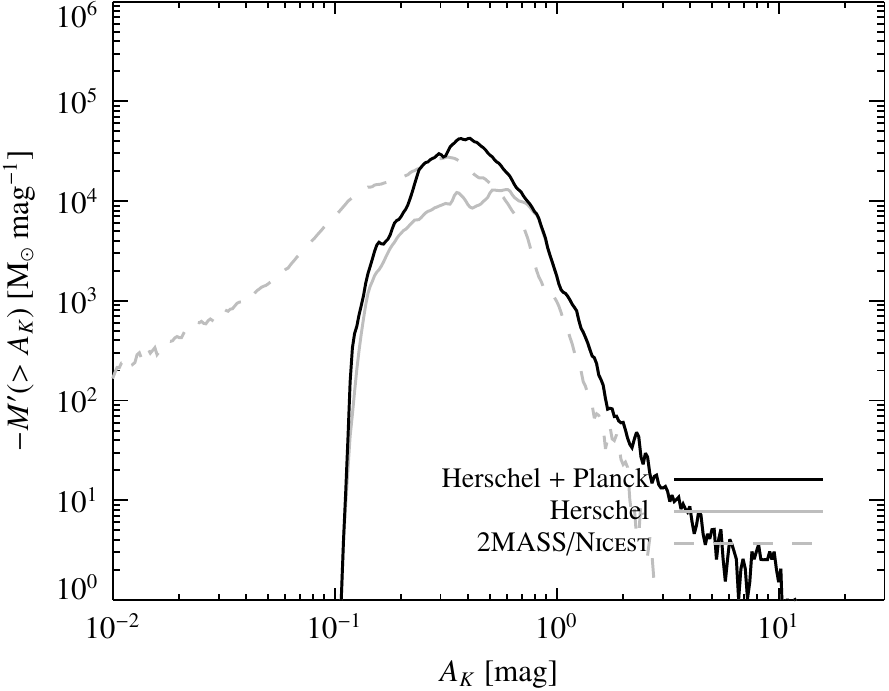}}
  \caption{Derivative of the integrated mass $M^{\prime}(>A_K)$ with respect to extinction as a function of extinction threshold. The colour coding of the lines is as in Fig.~\ref{fig:Sdist}.}
  \label{fig:Mprimedist}
\end{figure}

\end{appendix}

\end{document}